\documentclass[twocolumn,pra, aps,superscriptaddress]{revtex4}

\usepackage{mathptmx}
\usepackage{subfigure}
\usepackage{dcolumn}
\usepackage{amsmath,amssymb}
\usepackage{bm}
\usepackage{color}
\usepackage{latexsym}
\usepackage{epstopdf}
\usepackage{color}
\usepackage[english]{babel}
\usepackage{latexsym}
\usepackage{makeidx}

\usepackage{psfrag,graphicx} 
\usepackage{epsf} 
\usepackage{subfigure} 
\usepackage{amsmath} 
\usepackage{amssymb} 
\usepackage{amsfonts}
\usepackage{bm}
\usepackage{natbib}
\usepackage{epstopdf}
\usepackage{appendix}

\definecolor{mygrey}{gray}{0.35}
\definecolor{myblue}{rgb}{0.05,0.05,0.8}
\definecolor{myzard}{cmyk}{0,0,0.05,0}
\definecolor{mywhite}{rgb}{1,1,1}
\definecolor{mywhite}{rgb}{1,1,1}
\definecolor{myred}{rgb}{1,0.,0.3}

\usepackage[colorlinks=true,citecolor=myred,linkcolor=myblue]{hyperref}

\def\be{\begin{equation}}
\def\ee{\end{equation}}
\def\ba{\begin{align}}
\def\enda{\end{align}}
\def\bi{\begin{itemize}}
\def\ei{\end{itemize}}

 \def\ee{\mathord{\rm e}}
 
 \def\ii{\mathord{\rm i}}

\def\half{\textstyle\frac{1}{2}}

 \def\ee{\mathord{\rm e}}
 
 \def\ii{\mathord{\rm i}}

\def\half{\textstyle\frac{1}{2}}

\renewcommand{\ii}{{\rm i}}
\renewcommand{\ee}{{\rm e}}

\def\beq{\begin{equation}}
\def\beq{\begin{equation}}
\def\eeq{\end{equation}}

 \newcommand{\ket}[1]{|#1\rangle}
 \newcommand{\bra}[1]{\langle #1|}

\def\ii{{\bf i}}

\begin{document}


\title{Photon-Assisted-Tunneling Toolbox  for  Quantum Simulations in Ion Traps}

\author{Alejandro Bermudez}
\affiliation{
Institut f\"ur Theoretische Physik, Albert-Einstein Allee 11, Universit\"at Ulm, 89069 Ulm, Germany 
}

\author{Tobias Schaetz}
\affiliation{
Max-Planck-Institut f\"ur Quantenoptik, Hans-Kopfermann-Strasse 1, D-85748 Garching, Germany
}
\affiliation{Albert-Ludwigs-Universit\"at Freiburg, 
Physikalisches Institut, Hermann-Herder-Straße 3, 79104 Freiburg, Germany}
\author{Diego Porras}
\affiliation{
Departamento de F\'isica Te\'orica I,
Universidad Complutense, 
28040 Madrid, 
Spain
}


\begin{abstract}
We describe a versatile toolbox for the quantum simulation of many-body lattice models, capable of exploring the combined effects of background Abelian and non-Abelian gauge fields, bond and site disorder, and strong on-site interactions. We show how
 to control the quantum dynamics of particles trapped in lattice potentials by the photon-assisted tunneling induced by periodic drivings. This scheme is general enough to be applied to either bosons or fermions with the additional advantage of being non-perturbative. It finds an ideal application in microfabricated ion trap arrays, where the quantized vibrational modes of the ions can be described by a quantum lattice model. We present a detailed theoretical proposal for a quantum simulator in that experimental setup, and
show that it is possible to explore phases of matter that range from the fractional quantum Hall effect, to exotic strongly-correlated glasses, or  flux-lattice models decorated with arbitrary patterns of localized defects.  
\end{abstract}

\maketitle

\begingroup
\hypersetup{linkcolor=black}
\tableofcontents
\endgroup

\section{Introduction}

In general,  quantum many-body  systems  cannot be  understood by extrapolating the properties of their individual constituents~\cite{more}, but rather by considering their rich collective behavior. A prototypical example is that of sound waves in solids~\cite{feynman_lectures}, where the atoms do not move individually but vibrate collectively and give rise to propagating quasiparticles, the so-called  phonons. This particular many-body problem is exceptional since the properties of the phonons  can be calculated exactly, something that  rarely occurs in  quantum many-body systems. The usual paradigm is that the complexity of the model grows very fast with the number of particles, making both  analytical and numerical methods more involved and less efficient.  A radically different  approach is based on  the so-called {\it experimental quantum simulations}, which were envisaged several decades ago by R. P. Feynman~\cite{qs}, and have now evolved into a discipline that merges concepts from atomic physics, quantum optics, quantum-information science and condensed-matter physics. The central idea of a quantum simulation is to  manipulate  the microscopic properties of a particular experimental setup  in a way that it reproduces faithfully a quantum many-body model under study. In this way, nature itself computes the properties of the model, and our measurements yield the  answer to  questions such as the nature of the ground-state  and collective excitations, determining whether a dedicated model is sufficient to describe the relevant properties of a particular system. Accordingly, quantum simulations have the potential of solving  fundamental open problems in physics, ranging from high-temperature superconductivity, to the thermalization of closed quantum systems, or the properties of spin glasses. 

 Imagine for a moment that  the phonons in a solid were controllable to such an extent that their Hamiltonian could be engineered to  target a many-body model of interest. For instance, by shaping the crystal anharmonicities, one could tune the phonon-phonon interactions and reproduce the physics of strongly-correlated models.  Unfortunately, it is hard to develop  techniques that allow such microscopic control in solid-state materials, and it might be beneficial to search  for different experimental platforms. This exotic  idea may become realized  in  experiments with  Coulomb crystals of cold atomic ions in radio-frequency  traps~\cite{blatt_wineland}. Current experimental tools allow for a promising control of the vibrational excitations at the quantum level,  such that  phonon-based quantum simulations of many-body physics may become a reality~\cite{hubbard_porras,many_body_qs}.

So far, the most exploited property of the vibrational excitations of trapped ions  is their ability to  transfer information between distant ions,  which encode  quantum bits (qubits) in their electronic states. This provides a mechanism to perform quantum logic operations between distant qubits, a fundamental building block of quantum-information protocols~\cite{gates}. This type of two-qubit couplings can be interpreted as a spin-spin interaction where the qubit plays the role of a pseudospin (henceforth referred as spin)~\cite{ions_ising_interaction} and can be exploited for quantum-simulation purposes.  From this perspective,  the phonons act as mediators of the interaction and give rise to a wide range of spin models~\cite{ising_porras}, which are familiar in the field of quantum magnetism. The experimental success of these  spin-based quantum simulations, either in an analog~\cite{ising_exp} or digital version~\cite{digital}, has motivated a variety of proposals that range from neural networks~\cite{neural_networks_lewenstein}, to three-body interactions~\cite{3_spin_models}, frustrated magnetism~\cite{frustrated},  mixed-spin  models~\cite{impurity_magnetism}, or topologically-ordered spin models~\cite{top}.  

Instead of using the phonons as a gadget to obtain the desired quantum-spin simulator, one can substantially enhance the capabilities by reclaiming phonons as the building blocks for the quantum simulation of many-body  models~\cite{hubbard_porras}. The local vibrational excitations of each atomic ion, considered to be trapped individually, correspond to bosonic quasiparticles, and the Coulomb interaction is responsible for the interchange of these bosons  between distant ions. To build interesting bosonic quantum simulators, one must complement the aforementioned scheme with additional ingredients, such as strong trapping non-linearities that lead to Mott insulating~\cite{hubbard_porras} and frustrated phases~\cite {frustrated_hard_core_bosons_schmied}, or  incommensurate trapping potentials leading to different quantum phase transitions~\cite{kontorova_garcia}. Besides, the spins of the ions can  be used as a tool to widen the  applicability of this quantum simulator, which can potentially target the spin-boson model~\cite{porras_spin_boson},  the lattice Jaynes-Cummings model~\cite{lattice_JC_ivanov}, or the phenomenon of Anderson localization due to disorder~\cite{anderson}. In addition to these many-body quantum simulations, there has also been a recent activity in building  analogues of single-particle phenomena with a special emphasis on relativistic effects (see e.g.~\cite{ion_dirac}). In this case, even if the models are tractable on a classical computer,  the predicted effects  are hard  to access in the original experiments, and thus justify the effort in building a trapped-ion analogue. 

In this article, we introduce a versatile toolbox for the  quantum simulation of many-body lattice models. This toolbox is based on the phenomenon of  {\it photon-assisted tunneling} (PAT) of phonons in ion traps~\cite{gauge}, but can also be applied to different systems of bosons, and even fermions, in a lattice. As shown in this manuscript, the PAT effect has a   a variety of facets  that can be exploited to provide new paradigms for the aforementioned many-body quantum simulation. The idea underlying the PAT is that the tunneling of particles between the wells of a periodic lattice can be assisted by inducing resonances that correspond to the absorption/emission of photons out of an EM-field providing a periodic driving force. Even though this idea   was  initially introduced for condensed-matter systems~\cite{pat_cm} (see~\cite{pat_review} and references therein), it has also been applied in the field of ultracold  atoms in optical lattices~\cite{pat_ol}. Here, the analogy is straightforward since the neutral atoms can tunnel between the adjacent wells of a driven  periodic potential created by light. In this work, we show how to exploit the PAT in order to induce synthetic gauge fields in the aforementioned many-body lattice models.

In the context of trapped ions, the possibility of controlling the tunneling of the vibrational excitations between  ions stored in two separate potential minima, aligned within a linear radio-frequency trap, has been recently demonstrated~\cite{inhibitted_hopping_exp}. The exchange of phonons between two  ions   can be controlled and optimized by matching  their individual trapping frequencies. Building on these seminal experiments, we have demonstrated theoretically that driving the trapping frequencies  in/out of resonance realizes a novel version of the PAT paradigm~\cite{gauge}. Rather than relying on a periodic force~\cite{pat_cm,pat_ol} scaled to a larger number of ions, which may be quite demanding  in trapped-ion experiments, our scheme relies on the periodic modulation of the trapping frequencies. In particular, it requires an additional relative phase between the individual modulations that can be experimentally tuned. This allows for the full control of both the amplitude and the phase of the tunneling of phonons,  a result that becomes specially interesting for  two-dimensional arrays of micro-fabricated traps~\cite{microtraps}. Here, the technology of ion-trap micro-fabrication provides a means of assembling arrays of ion traps in any desired geometry~\cite{schmied_micro}. The combination of the capabilities for the design and fabrication of arrays of microtraps with the tool of PAT~\cite{gauge} opens a vast amount of possibilities for a quantum simulator of many-body physics. These range from the realization of bosonic models subjected to {\it Abelian and non-Abelian synthetic gauge field}s, to bond and site disorder leading to {\it glassy phases}, or to decorated flux lattices which can be related to {\it anyonic excitations}. In this article, we provide a detailed analysis of such PAT in arrays of microtraps.

This article is organized as follows. The main results  are summarized in Sec.~\ref{summary}, so that this section can be consulted  for a general overview. The general scheme of PAT for any lattice model is introduced in Sec.~\ref{pat_general}, where we demonstrate how the  synthetic gauge field arises from the assisted tunneling. The application as a toolbox for quantum simulations with trapped ions is contained in Sec.~\ref{pat_ions}. Here, we present a thorough description of the different many-body models that can be explored by exploiting the peculiarities of trapped ions, which widen the versatility of our quantum simulator.

\section{Photon-assisted tunneling and synthetic gauge fields}
\label{pat_general}
In this Section, we  describe the concept of PAT for a general system, and discuss how it can be exploited for the quantum simulation of lattice models incorporating synthetic gauge fields. Once the main ingredients are identified, we focus on the case of ions in micro-fabricated traps in Sec.~\ref{pat_ions}, and show how to implement the PAT for the vibrational excitations. Let us remark that the generality of this section may also find applications in different platforms, such as photons in arrays of cavities in circuit QED~\cite{photon_cavities}, or ultracold atoms in optical lattices~\cite{pat_greiner}. We also note that  the PAT scheme is not restricted to bosons, but works equally well for fermions. This may alleviate some of the difficulties that arise in  the simulation of synthetic gauge fields via Raman-assisted tunneling of fermionic atoms  in spin-dependent optical lattices~\cite{laser_assisted_ol}. 

We consider a tight-binding model  describing the tunneling of particles, either bosons or fermions, between the sites of an underlying two-dimensional lattice.  The lattice sites are characterized by  a vector of integer numbers ${\bf i}=(i_1,i_2)$, such that ${\bf r}_{\ii}^0=i_1d_1{\bf e}_1+i_2d_2{\bf e}_2$, where $\{{\bf e}_{\alpha}\}$ are the unit vectors  spanning the lattice, and  $\{d_{\alpha}\}$ are the corresponding lattice constants. The particles are represented by bosonic or fermionic creation-annihilation operators $a_{\sigma,\bf i}^{\dagger},a_{\sigma,\bf i}^{\phantom{\dagger}}$ where $\sigma$ labels some additional degrees of freedom. The dynamics of the system is described by the following Hamiltonian
\begin{equation}
\label{tight_binding_general}
H=H_0+H_{\rm t}=\sum_{\sigma,{\bf i}}\omega_{\sigma,{\bf i}}a_{\sigma,\bf i}^{\dagger}a_{\sigma,\bf i}^{\phantom{\dagger}}+\sum_{\sigma}\sum_{{\bf i}>{\bf j}}\big(J_{{\rm t};{\bf i}{\bf j}}^{\sigma}a_{\sigma,\bf i}^{\dagger}a_{\sigma,\bf j}^{\phantom{\dagger}}+\text{H.c.}\big),
\end{equation} 
where $\omega_{\sigma,{\bf i}}$ stands for the on-site energy ($\hbar=1$), and $J_{{\rm t}; {\bf i}{\bf j}}^{\sigma}$ is the tunneling amplitude of the particles between different lattice sites ${\bf j}\to{\bf i}$, which usually depends on the  distance between sites such that $J_{{\rm t}; {\bf i}{\bf j}}^{\sigma}=J_{{\rm t}}^{\sigma}(|{\bf r}_{\bf i }^0-{\bf r}^0_{\bf j}|)$. The tunneling constraint ${\bf i>{\bf j}}$ refers 
to an ordering of the sites, such that $i_1 > j_1$, or $i_2 > j_2$ if $i_1 = j_1$.
Two additional ingredients are required:

\begin{figure}
\centering
\includegraphics[width=1\columnwidth]{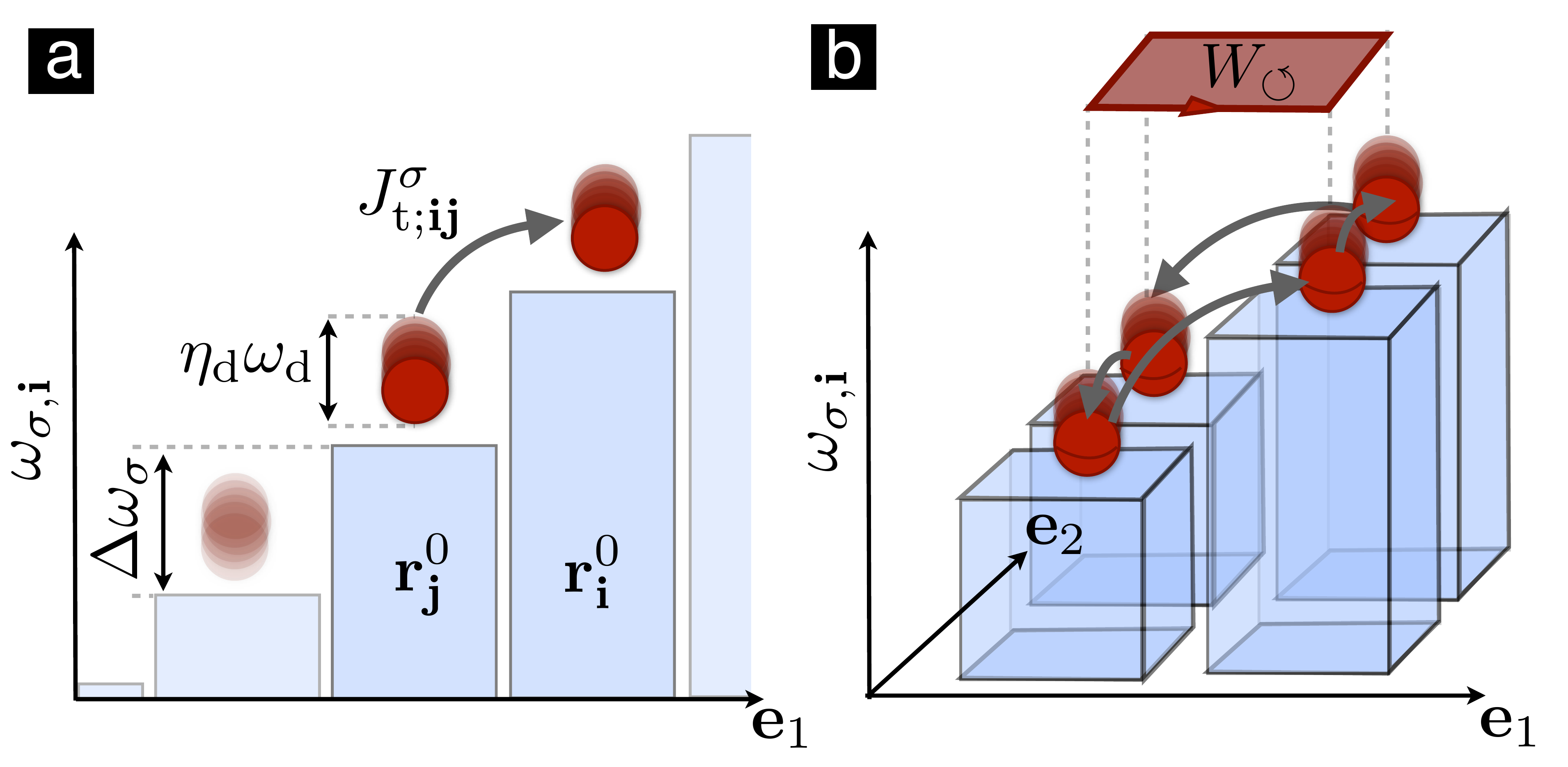}
\caption{ {\bf Photon-assisted tunneling scheme:} {\bf (a)} Scheme for the driven one-dimensional tight-binding model in Eqs.~\eqref{tight_binding_general} and~\eqref{driving}. The tunneling of the particles between neighboring sites ${\bf r}_{\bf j}^0\to{\bf r}_{\bf i}^0$, which is initially suppressed by the large gradient $\Delta\omega_{\sigma}\gg J_{\rm t;{\bf ij}}^{\sigma}$, becomes assisted by a resonant periodic modulation of the on-site energies with amplitude $\eta_{\rm d}\omega_{\rm d}$. {\bf (b)} Two-dimensional scheme giving rise to a synthetic gauge field such that the path around an elementary plaquette $W_{\circlearrowleft}\propto\ee^{i\phi_{\circlearrowleft}}$ mimics the Aharonov-Bohm phase, which is picked up by a charged particle following the path around a plaquette pierced by  an external magnetic field orthogonal to the lattice.}
\label{fig_pat_scheme}
\end{figure}

{\it i) Gradient of the on-site energies:} The on-site energies  have the following expression $\omega_{\sigma,{\bf i}}= \omega_{\sigma}+\Delta\omega_{\sigma} i_1$. Here, $\omega_{\sigma}$ is a constant energy offset, and $\Delta\omega_{\sigma}$ results due to a gradient along one of the lattice principal axes satisfying 
\begin{equation}
\label{gradient_constraint}
J_{{\rm t;}\bf ij}^{\sigma}\ll \Delta\omega_{\sigma}.
\end{equation}

{\it ii) Periodic modulation of the on-site energies:} The on-site energies must be supplemented by the   periodic modulation $\omega_{\sigma,{\bf i}}\to \omega_{\sigma,{\bf i}}+\eta_{\rm d}\omega_{\rm d}\cos (\omega_{\rm d}t+\phi_{\bf i})$, where $\omega_{\rm d}$ is the driving frequency, and $\eta_{\rm d}\omega_{\rm d}$ the driving amplitude. Note that the periodic driving incorporates a site-dependent phase $\phi_{\bf i}$ that shall play a crucial role for the PAT, and is  described by
\begin{equation}
\label{driving_constraint}
\phi_{\bf i}=\phi_1i_1+\phi_2i_2.
\end{equation}

These two ingredients, schematically represented in Fig.~\ref{fig_pat_scheme}{\bf (a)}, are incorporated in the above description by modifying the Hamiltonian $H_0\to H_0(t)$ as follows
\begin{equation}
\label{driving}
H_0(t)=\sum_{\sigma,\bf i}\big(\omega_{\sigma}+\Delta\omega_{\sigma} i_1+\eta_{\rm d}\omega_{\rm d}\cos (\omega_{\rm d}t+\phi_{\bf i})\big)a_{\sigma,{\bf i}}^{\dagger}a_{\sigma,{\bf i}}^{\phantom{\dagger}}.
\end{equation}
Let us emphasize again that the standard formulation of the PAT relies on a periodic force acting on the particles residing on the lattice sites~\cite{pat_cm,pat_ol}. In contrast, our scheme considers a periodic modulation of the on-site energies, which turns out to be better suited to taylor the tunneling amplitudes in analogy with a  background synthetic gauge field. In subsection~\ref{analytical}, we present an analytical model, which is confronted to numerical simulations in~\ref{numerical}.

\subsection{Analytical description}
\label{analytical}

\subsubsection{ Dressed tunneling} 
 In this subsection, we derive a compact expression for the PAT strength, and study its  dependence on the periodic-driving parameters. Let us express the tunneling Hamiltonian $H_{\rm t}$ in the interaction picture with respect to $H_0(t)$, namely $H_{\rm t}(t)=U(t)^{\dagger}H_{\rm t}U(t)$ where $U(t)=\ee^{-i\int_0^t{\rm d}\tau H_0(\tau)}$. In this picture, the  annihilation operators fulfill 
\begin{equation}
i\frac{{\rm d}a_{\sigma,{\bf i}}}{{\rm d}t}=( \omega_{\sigma}+\Delta\omega_{\sigma} i_1+\eta_{\rm d}\omega_{\rm d}\cos (\omega_{\rm d}t+\phi_{\bf i}))a_{\sigma,{\bf i}},
\end{equation}
which leads to the following relation between  both pictures
\begin{equation}
a_{\sigma,{\bf i}}(t)=\ee^{-i(\omega_{\sigma}+\Delta\omega i_1)t}\ee^{-i\eta_{\rm d}\sin(\omega_{\rm d}t+\phi_{\bf i})}\ee^{+i\eta_{\rm d}\sin(\phi_{\bf i})}a_{\sigma,{\bf i}}.
\end{equation}
At this point, we note that the  tunneling Hamiltonian $H_{\rm t}$ is invariant under U(1) gauge transformations, and thus,  the last term in the above expression can be trivially gauged away $a_{\sigma,{\bf i}}\to\ee^{-i\eta_{\rm d}\sin(\phi_{\bf i})}a_{\sigma,{\bf i}}$. Accordingly, the Hamiltonian becomes  $H_{\rm t}(t)=\sum_{{\bf i}>{\bf j}}J_{{\rm d};{\bf i}{\bf j}}^{\sigma}(t)a_{\sigma,{\bf i}}^{\dagger}a_{\sigma,{\bf j}}^{\phantom{\dagger}}+\text{H.c.}$, where $J_{{\rm d};{\bf i}{\bf j}}^{\sigma}(t)=J_{{\rm t};{\bf i}{\bf j}}^{\sigma}\Theta(t)$, and the time-dependence is encoded in the  function 
\begin{equation}
\label{total_hopping}
\Theta(t)=\left\{\begin{array}{c} \ee^{i\eta_{\rm d}\big(\sin(\omega_{\rm d}t+\phi_{\bf i})-\sin(\omega_{\rm d}t+\phi_{\bf j})\big)}\ee^{i\Delta\omega_{\sigma}(i_1-j_1)t},\hspace{1ex} \text{if} \hspace{1ex} i_1>j_1
 \\ \ee^{i\eta_{\rm d}\big(\sin(\omega_{\rm d}t+\phi_{\bf i})-\sin(\omega_{\rm d}t+\phi_{\bf j})\big)}\phantom{\ee^{i\Delta\omega_{\sigma}(i_1-j_1)t}},
\hspace{1ex} \text{if} \hspace{0.7ex} i_1=j_1.
\end{array}\right.
\end{equation}
In this expression, one readily observes that the tunneling amplitude  becomes dressed by the photons of the  periodic driving $J_{{\rm t};{\bf i}{\bf j}}^{\sigma}\to J_{{\rm d};{\bf i}{\bf j}}^{\sigma}(t)$. Besides, the fundamental role of the phase of the periodic driving also becomes apparent:  only when $\phi_{\bf i}\neq\phi_{\bf j}$,    the tunneling becomes assisted $J_{{\rm d};{\bf i}{\bf j}}^{\sigma}(t)\neq J_{{\rm t};{\bf i}{\bf j}}^{\sigma}$.   In order to proceed with this analytical treatment, we  use the   identity
\begin{equation}
\ee^{i\eta_{\rm d}\sin(\omega_{\rm d}t+\phi_{\bf i})}=\sum_{s\in\mathbb{Z}}J_s(\eta_{\rm d})\ee^{is(\omega_{\rm d}t+\phi_{\rm i})},
\end{equation}
where $J_s(\eta_{\rm d})$ are Bessel functions of the first class~\cite{ab_st}. Hence, the dressed tunneling  becomes a sum of terms with different time dependences. Let us first  focus on the tunneling along the gradient, $i_1>j_1$, which can be expressed as follows
\begin{equation}
J_{{\rm d};{\bf i}{\bf j}}^{\sigma}(t)=J_{{\rm t};{\bf i}{\bf j}}^{\sigma}\hspace{-0.75ex}\sum_{s,s'\in\mathbb{Z}}\hspace{-0.75ex}J_s(\eta_{\rm d})J_{s'}(\eta_{\rm d})\ee^{i\Delta\omega_{\sigma}(i_1-j_1)t}\ee^{i\big((s-s')\omega_{\rm d}t+s\phi_{\bf i}-s'\phi_{\bf j})\big)}.
\end{equation}
 By tuning the driving frequency to $\omega_{\rm d}=\Delta\omega_{\sigma}/r$, where $r$ is some positive integer, the above expression contains two different types of terms. There are  resonant terms  fulfilling
  \begin{equation}
  s'=s+r(i_1-j_1),  
  \end{equation}
   whereas the remaining terms are far off-resonance. By applying  a rotating-wave approximation (RWA) for $J_{{\rm t};{\bf i}{\bf j}}^{\sigma}\ll\Delta\omega_{\sigma}$~\eqref{gradient_constraint}, we neglect  the rapidly-oscillating terms   yielding
   \begin{equation}
J_{{\rm d};{\bf i}{\bf j}}^{\sigma}=J_{{\rm t};{\bf i}{\bf j}}^{\sigma}\hspace{-0.ex}\sum_{s\in\mathbb{Z}}\hspace{-0.ex}J_s(\eta_{\rm d})J_{s+r(i_1-j_1)}(\eta_{\rm d})\ee^{i \big(s\phi_{\bf i}-(s+r(i_1-j_1))\phi_{\bf j}\big)}.
\end{equation}
 According to this expression, both the amplitude and the phase of the tunneling are controlled by the periodic driving parameters.  As shown in Sec.~\ref{pat_ions}, the periodic driving can be achieved by means of optical dipole forces on the ions, and  the tunneling along the ${\bf e}_1$-axis is thus assisted by the absorption/emission of $r(i_1-j_1)$ photons from the EM-field providing the  driving force, hence the name PAT. A similar analysis for the tunneling orthogonal to the gradient, $i_1=j_1$, shows that the  resonant terms satisfy $s'=s$. 
 
 The complete assisted-tunneling Hamiltonian becomes
 \begin{equation}
 \label{assisted_tunneling_hamiltonian}
 H_{\rm eff}=\sum_{\sigma}\sum_{{\bf i}>{\bf j}}J_{{\rm d};{\bf i}{\bf j}}^{\sigma}a_{\sigma,{\bf i}}^{\dagger}a_{\sigma,{\bf j}}^{\phantom{\dagger}}+\text{H.c.},
 \end{equation}
 with the dressed couplings  expressed as follows
    \begin{equation}
   \label{dressed_coupling}
J_{{\rm d};{\bf i}{\bf j}}^{\sigma}=  J_{{\rm t};{\bf i}{\bf j}}^{\sigma}\mathcal{F}_{f({\bf i},{\bf j})}(\eta_{\rm d},\eta_{\rm d},\Delta\phi_{{\bf i}{\bf j}})\ee^{-i\frac{f({\bf i},{\bf j})}{2}(\phi_{{\bf i}}+\phi_{\bf j})},
   \end{equation}
  where the function $\mathcal{F}$  is responsible for the modulation of the tunneling amplitude
  \begin{equation}
  \mathcal{F}_{\chi}(\zeta,\xi,\theta)=\sum_{s\in\mathbb{Z}}J_s(\zeta)J_{s+\chi}(\xi)\ee^{i(s+\frac{\chi}{2})\theta},
  \end{equation}
  and we have defined the phase difference $\Delta\phi_{{\bf i}{\bf j}}=\phi_{\bf i}-\phi_{\bf j}$, and the function $f({\bf i},{\bf j})=r(i_1-j_1)$.
  
\begin{figure}
\centering
\includegraphics[width=1\columnwidth]{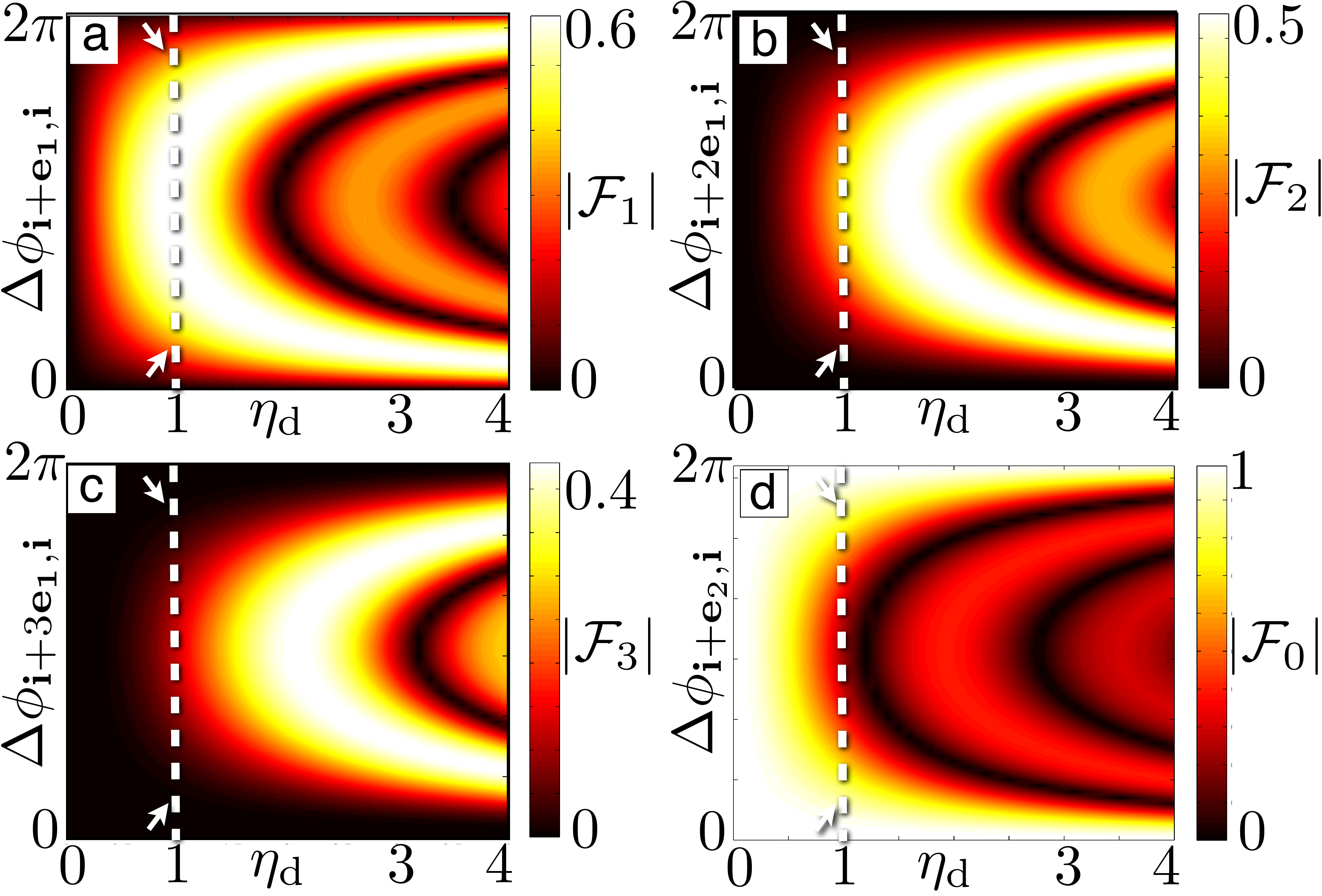}
\caption{ {\bf Photon-assisted modulation of the tunneling amplitude:}  Contour plot of the modulation amplitude $|\mathcal{F}_{r(i_1-j_1)}|$ for the tunneling between sites ${\bf i}\to{\bf j}$, such that $i_1>j_1$, as a function of the driving parameters $\eta_{\rm d},\Delta\phi_{{\bf i}{\bf j}}$ for $r=1$. {\bf (a)} First-neighbor assisted tunneling (${\bf i} \to {\bf i} + {\bf e}_1$), {\bf (b)} Second-neighbor assisted tunneling (${\bf i} \to {\bf i} + 2 {\bf e}_1$), {\bf (c)} Third-neighbor assisted tunneling (${\bf i} \to {\bf i} + 3 {\bf e}_1$), {\bf (d)} First neighbor assisted tunneling orthogonal to the gradient ${\bf i} \to {\bf i} + {\bf e}_2$.}
\label{fig_amplitude}
\end{figure}

Let us  emphasize one of the important properties of this photon-assisted scheme, namely, its non-perturbative character. So far, we have not used any restriction on the values of $\eta_{\rm d}$ or $\phi_{\bf i}$, so that  the dressed tunneling $J_{\rm d}$ may be of the same order as the bare one $J_{\rm t}$. In Fig.~\ref{fig_amplitude}, we represent the modulation function $\mathcal{F}_{f({\bf i},{\bf j})}(\eta_{\rm d},\eta_{\rm d},\Delta\phi_{{\bf i}{\bf j}})$ for different periodic-driving parameters, and different ranges of the tunneling. In particular, Figs.~\ref{fig_amplitude}{\bf (a)-(c)} represent the tunneling along the direction of the gradient  for $r=1$, whereas Fig.~\ref{fig_amplitude}{\bf (d)} stands for the tunneling orthogonal to the gradient. As derived from Fig.~\ref{fig_amplitude}{\bf (a)}, for $\eta_{\rm d}\approx 1,\Delta\phi=\pi$, one gets the maximal amplitude for the assisted tunneling between  nearest neighbors. In particular, we find $|\mathcal{F}_1|\approx 0.6$ for these optimal values, which thus supports our previous claim about the non-perturbative nature of the scheme (i.e. $|J^{\sigma}_{{\rm d};{\bf i}+{\bf e}_1,{\bf i}}|\approx 0.6|J^{\sigma}_{{\rm t};{\bf i}+{\bf e}_1,{\bf i}}|$). It is also interesting to note that, as we consider longer range terms, their amplitude gets gradually diminished, as shown in Figs.~\ref{fig_amplitude}{\bf (b)-(c)}. In  Fig.~\ref{fig_amplitude}{\bf (d)}, we represent the modulation amplitude in the direction perpendicular to the gradient, which taking the same parameters as above leads to $|J^{\sigma}_{{\rm d};{\bf i}+{\bf e}_2,{\bf i}}|\approx 0.2|J^{\sigma}_{{\rm t};{\bf i}+{\bf e}_2,{\bf i}}|$.

Before concluding this subsection, we briefly comment on another interesting property of this scheme. Let us consider that the original Hamiltonian~\eqref{tight_binding_general} also contains on-site particle-particle interactions $H+V$, where
\begin{equation}
\label{onsite_interactions}
V=\sum_{\sigma\sigma'}\sum_{{\bf i}}U_{\sigma\sigma'}a_{\sigma,{\bf i}}^{\dagger}a_{\sigma',{\bf i}}^{\dagger}a_{\sigma',{\bf i}}^{\phantom{\dagger}}a_{\sigma,{\bf i}}^{\phantom{\dagger}},
\end{equation}
and the dependence of the interaction strengths $U_{\sigma\sigma'}$ on the internal indices depends on the bosonic/fermionic nature of the particles. It is  straightforward to see that the gradient and periodic driving  do not modify $V$. Therefore, the PAT also works in the presence of on-site interactions. It would  be interesting to study how the scheme can be used to modify long-range interactions, but this lies beyond the scope of the present work.
   
\subsubsection{ Synthetic gauge fields}
In this subsection, we demonstrate that it is possible to interpret the phase of the dressed tunneling~\eqref{dressed_coupling}  as if it was originated by a background gauge field.  In particular, we show that whenever particles tunnel along a closed path in the lattice, they pick up a non-vanishing phase  analogous to the celebrated Aharonov-Bohm phase for charged particles in electromagnetic fields~\cite{ab_phase}.  The consecutive  tunneling of a particle  around a unit plaquette of the lattice ${\bf i}\to{\bf i}+{\bf e}_1\to{\bf i}+{\bf e}_1+{\bf e}_2\to {\bf i}+{\bf e}_2\to{\bf i}$ [Fig.~\ref{fig_pat_scheme}{\bf (b)}]  is formally given by 
\begin{equation}
W^{(1)}_{\circlearrowleft}=J^{\sigma}_{{\rm d};{\bf i},{\bf i}+{\bf e}_2}J^{\sigma}_{{\rm d};{\bf i}+{\bf e}_2,{\bf i}+{\bf e}_1+{\bf e}_2}J^{\sigma}_{{\rm d};{\bf i}+{\bf e}_1+{\bf e}_2,{\bf i}+{\bf e}_1}J^{\sigma}_{{\rm d};{\bf i}+{\bf e}_1,{\bf i}}.
\end{equation}
Using Eqs.~\eqref{total_hopping} and~\eqref{dressed_coupling}, it can be expressed as
\begin{equation}
W^{(1)}_{\circlearrowleft}=|J^{\sigma}_{{\rm t}}(d_2)J^{\sigma}_{{\rm t}}(d_1)\mathcal{F}_0(\eta_{\rm d},\eta_{\rm d},\phi_2)\mathcal{F}_r(\eta_{\rm d},\eta_{\rm d},\phi_1)|^2\ee^{ir\phi_2},
\end{equation}
which leads to $W^{(1)}_{\circlearrowleft}=|W^{(1)}_{\circlearrowleft}|\ee^{i\phi_{\circlearrowleft}}$, where $\phi_{\circlearrowleft}=r\phi_2$ only depends on the component of the periodic-driving phase that is orthogonal to the gradient. This quantity is a gauge-invariant observable proportional to the so-called Wilson loop in lattice gauge theories~\cite{lgt}, which can be expressed as follows
\begin{equation}
\label{ab_phase}
W^{(1)}_{\circlearrowleft}\propto\ee^{ie^*\oint_{\circlearrowleft}{\rm d}{\bf r}\cdot{\bf A}_{\rm s}}=\ee^{ie^*\int_{\square}{\bf B}_{\rm s}\cdot {\rm d}{\bf S} }.
\end{equation}
Here, we have introduced an effective charge $e^*$  independent of the charged/neutral character of the particles, the synthetic gauge potential ${\bf A}_{\rm s}$, and the synthetic gauge field ${\bf B}_{\rm s}$.   Independently on how we rearrange  the phases of the tunneling strengths locally, the enclosed phase $\phi_{\circlearrowleft}=e^*\int_{\square}{\bf B}_{\rm s}\cdot {\rm d}{\bf S}$ shall always be left invariant, and can be thus interpreted as the magnetic flux of a synthetic magnetic  field  ${\bf B}_{\rm s}$ that pierces the lattice.  In order for this analogy to be complete, one should consider carefully the long-range character of the tunneling, according to which particles can follow different closed paths. We shall explore the two possible paths around the second smallest plaquettes, namely 
\begin{equation}
\begin{split}
W^{(2)}_{\circlearrowleft}=J^{\sigma}_{{\rm d};{\bf i},{\bf i}+2{\bf e}_2}J^{\sigma}_{{\rm d};{\bf i}+2{\bf e}_2,{\bf i}+{\bf e}_1+2{\bf e}_2}J^{\sigma}_{{\rm d};{\bf i}+{\bf e}_1+2{\bf e}_2,{\bf i}+{\bf e}_1}J^{\sigma}_{{\rm d};{\bf i}+{\bf e}_1,{\bf i}},\\
W'^{(2)}_{\circlearrowleft}=J^{\sigma}_{{\rm d};{\bf i},{\bf i}+{\bf e}_2}J^{\sigma}_{{\rm d};{\bf i}+{\bf e}_2,{\bf i}+2{\bf e}_1+{\bf e}_2}J^{\sigma}_{{\rm d};{\bf i}+2{\bf e}_1+{\bf e}_2,{\bf i}+2{\bf e}_1}J^{\sigma}_{{\rm d};{\bf i}+2{\bf e}_1,{\bf i}}.\\
\end{split}
\label{flux.2}
\end{equation}
Repeating the above calculations, we  find that $W^{(2)}_{\circlearrowleft}/|W^{(2)}_{\circlearrowleft}|=W'^{(2)}_{\circlearrowleft}/|W'^{(2)}_{\circlearrowleft}|=\ee^{i2\phi_{\circlearrowleft}}$. Hence, the Aharonov-Bohm phase is doubled with respect to the unit plaquette, which  is consistent with the fact that the enclosed area is also doubled for these paths. The same occurs for any other closed path, and thus  the analogy to a background gauge field is consistent with the possible long-range of the tunnelings. 

At this point, it is worth emphasizing the generality of the scheme hereby proposed. It works  for both bosons and fermions, for any range of the tunneling  $J_{{\rm t};{\bf i}{\bf j}}^{\sigma}=J_{{\rm t}}^{\sigma}(|\bf{r}_{\bf i}^0-\bf{r}_{\bf j}^0|)$, it can incorporate local particle-particle interactions, and it is non-perturbative. In the second part of this manuscript, we shall specify this scheme to trapped-ion experiments, which provide an ideal platform where to realize a bosonic PAT (i.e. vibrational excitations). Before moving onto the numerical verification of these results, let us introduce an alternative and more compact formulation of the synthetic gauge fields. It is possible to rewrite the PAT Hamiltonian in analogy to the so-called Peierls substitution~\cite{peierls}, which yields
\begin{equation}
\label{peierls}
H_{\rm eff}=\sum_{\sigma}\sum_{{\bf i}>{\bf j}}\tilde{J}_{{\rm d};{\bf i}{\bf j}}^{\sigma}\ee^{ie^*\int_{{\bf j}}^{{\bf i}}{\rm d}{\bf r}\cdot {\bf A}_{\rm s}}a_{\sigma,{\bf i}}^{\dagger}a_{\sigma,{\bf j}}^{\phantom{\dagger}}+\text{H.c.}, 
\end{equation}
where we have introduced the dressed-tunneling amplitude $\tilde{J}_{{\rm d};{\bf i}{\bf j}}^{\sigma}=J_{{\rm t};{\bf i}{\bf j}}^{\sigma}\mathcal{F}_{f({\bf i},{\bf j})}(\eta_{\rm d},\eta_{\rm d},\Delta\phi_{{\bf i}{\bf j}})$, and  the synthetic gauge potential  ${\bf A}_{\rm s}({\bf x})=-B_0y{\bf e}_1$, where $B_0=r\phi_2/e^*d_1d_2$. This gauge potential, which corresponds to the famous Landau gauge for electrons in a constant magnetic field, is obtained after the following U(1) gauge transformation $a_{\sigma,{\bf i}}\to a_{\sigma,{\bf i}}\ee^{-i\chi_{\bf i}}$, where $\chi_{\bf i}=\phi_1i_1^2/2$. We note that this transformation is also consistent with the long-range character of the tunneling. This compact formulation~\eqref{peierls} condenses the main result of this section: a gradient and a periodic modulation of the on-site energies give us full access to the amplitude and phase of the tunneling, such that the effects of background synthetic gauge fields can be mimicked even for neutral particles or quasiparticles. 

\subsection{Numerical simulations}
\label{numerical}
In this subsection, we confront the analytical model for the PAT in Eq.~\eqref{peierls} with numerical simulations for the tunneling Hamiltonian~\eqref{tight_binding_general} subjected to the gradient and periodic driving~\eqref{driving}. We analyze in detail some basic realizations of the PAT, which allow us to carry out a thorough numerical study considering even the effects of finite temperatures.

\subsubsection{Photon-assisted tunneling along one link}
\begin{figure}
\centering
\includegraphics[width=1\columnwidth]{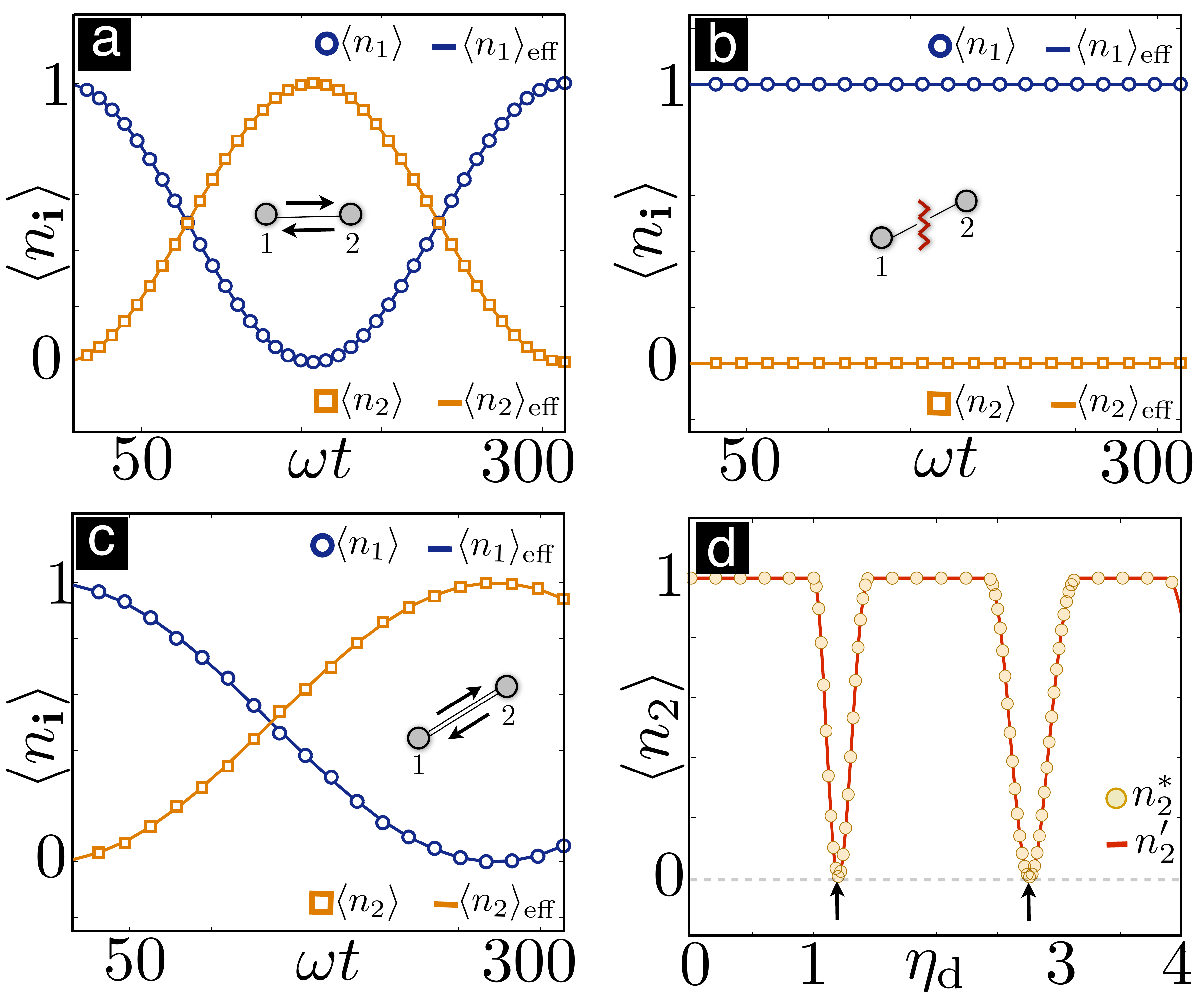}
\caption{ {\bf Photon-assisted tunneling along a link:} {\bf (a)} Bare tunneling for an initial state $\ket{\Psi_0}=a_1^{\dagger}\ket{\rm vac}$ with a single particle occupying the first site. The numerical expectation values $\langle n_1(t)\rangle$ (blue circle), and $\langle n_2(t)\rangle$ (orange squares) are compared to the effective description $\langle n_1(t)\rangle_{\rm eff}$ (blue lines), and $\langle n_2(t)\rangle_{\rm eff}$ (orange lines) described in Eq.~\eqref{exact}. {\bf (b)} Suppressed tunneling due to the gradient, and in the absence of periodic driving. {\bf (c)} Assisted tunneling in the presence of a periodic driving for $\Delta\phi_{12}=\pi$. {\bf (d)}  Coherent destruction of tunneling (black arrows) due to a periodic driving in the absence of the gradient. The yellow circles represent the  maximum population transferred to site 2 $n_2^*={\rm max}\{\langle n_2(t)\rangle: t\leq t^*\}$, such that $t^*=100\pi/|\mathcal{F}_{0}(\eta_{\rm d},\pi)|\omega$, as obtained from the numerical integration of Eqs.~\eqref{tight_binding_general} and~\eqref{driving}. The red line represents the predictions for the same magnitude $n_2'$ according to the effective Hamiltonian~\eqref{assisted_tunneling_hamiltonian}.  }
\label{fig_line}
\end{figure}
 Let us initially focus on the simplest situation to test the accuracy of the analytical model, namely, a lattice consisting of only two sites populated by spinless bosons. We consider the following parameters of  the driven Hamiltonian in Eqs.~\eqref{tight_binding_general} and~\eqref{driving}: the bare tunneling is $J_{{\rm t};12}=10^{-2}\omega$,  the gradient $\Delta\omega=0.5\omega$, and the periodic driving parameters correspond to $\eta_{\rm d}=1$, and $\omega_{\rm d}=\Delta\omega$. Note that we do not consider additional degrees of freedom, avoiding thus the index $\sigma$ in the following. In Fig.~\ref{fig_line}, we represent the expectation values $\langle n_i(t)\rangle=\bra{\Psi(t)} a_i^{\dagger}a_i\ket{\Psi(t)}$ that result from the numerical integration of the Schr\"{o}dinger equation $i{\rm d}\ket{\Psi(t)}/{\rm d}t=(H_0(t)+H_{\rm t})\ket{\Psi(t)}$, and compare them with the effective analytical description $\ket{\Psi(t)}_{\rm eff}=\ee^{-i H_{\rm eff}t}\ket{\Psi_0}$. We consider  an initial state with a single bosonic particle in the first site $\ket{\Psi_0}=a_1^{\dagger}\ket{\rm vac}$, where $\ket{\rm vac}$ stands for the vacuum. In this case, the effective description~\eqref{assisted_tunneling_hamiltonian} can be integrated exactly, yielding the following periodic oscillations of the particle between the two lattice sites 
\begin{equation}
\label{exact}
\begin{split}
\langle n_1(t)\rangle_{\rm eff}&=\half(1+\cos\omega_{\text{eff}}t),\\
\langle n_2(t)\rangle_{\rm eff}&=\half(1-\cos\omega_{\text{eff}}t),
\end{split}
\end{equation}
where $\omega_{\text{eff}}=2|J_{\rm d;12}|$ is the frequency of oscillations. In the absence of any gradient or periodic driving, $\omega_{\text{eff}}=2|J_{\rm t;12}|=0.02\omega$, and thus the particle exchange undergoes oscillations with a period of $T_{{\rm t};12}=\pi/|J_{{\rm t};12}|=100\pi/\omega$, which coincides exactly with the scale shown in Fig.~\ref{fig_line}{\bf (a)}. In this figure, we reveal that the effective description (circles and squares) agrees with the  numerical results (solid lines). Note that in order to carry out the numerical integration, we truncate the bosonic Hilbert space to $n_{\rm trunc}=4$. We have confirmed that the truncation  error after changing  $n_{\rm trunc}\to \tilde{n}_{\rm trunc}=n_{\rm trunc}+1$ lies below $|\langle n_i(t)\rangle_{n_{\rm trunc}}-\langle n_i(t)\rangle_{\tilde{n}_{\rm trunc}} |< 10^{-4}$. 
The same occurs in Fig.~\ref{fig_line}{\bf (b)}, where we  switch on the gradient. As a consequence of the mismatch between the on-site energies, the tunneling is completely inhibited. In order to assist it, we switch on the periodic driving in Fig.~\ref{fig_line}{\bf (c)} for $\phi_1=\phi_2+\pi$, which activates again the periodic oscillations but with a different period $T_{{\rm t};12}=\pi/|J_{{\rm d};12}|=100\pi/|\mathcal{F}_{1}(\eta_{\rm d},\eta_{\rm d},\pi)|\omega$. For the parameters used, we obtain $|\mathcal{F}_{1}(\eta_{\rm d},\eta_{\rm d},\pi)|\approx0.6$ (Fig.~\ref{fig_amplitude}{\bf (a)}), which explains the slightly longer oscillation period. However, we remark that both dynamics occur on the same time-scale as a consequence of the non-perturbative character of the scheme.

In Fig.~\ref{fig_line}{\bf (d)}, we study the maximal population transfer to site 2  due to the periodic driving $\eta_{\rm d}\neq0$, but in the absence of the gradient $\Delta\omega=0$.  We set the phases to  $\phi_1=0,\phi_2=\pi$, and study  the maximal transfer $n_2^*={\rm max}\{\langle n_2(t)\rangle: t\leq t^*\}$, such that $t^*=100\pi/|\mathcal{F}_{0}(\eta_{\rm d},\eta_{\rm d},\pi)|\omega$ for a range of driving strengths $\eta_{\rm d}\in[0,4]$. As shown in this figure, there are certain values of the driving strength, marked by black arrows, where the tunneling gets completely suppressed, a phenomenon  known as coherent destruction of tunneling~\cite{pat_cm}, or dynamic localization.
Let us remark that  this scheme leads to a perfect localization of the particles also in a longer one-dimensional chain, since the  dressed tunneling cancels for all  pairs of nearest-neighboring lattice sites simultaneously.  Once we set $\phi_{i_1}=\pi i_1/2$, such that $\Delta\phi_{i_1i_1+1}=\pi$,  it is possible to find a zero of the modulating function  $\mathcal{F}_0(\eta_{\rm d}^0,\eta_{\rm d}^0,\pm \pi)=0$ which is related to the particular values of the Bessel functions. In this limit,  the particles shall not diffuse through the lattice.

\begin{figure}
\centering
\includegraphics[width=1\columnwidth]{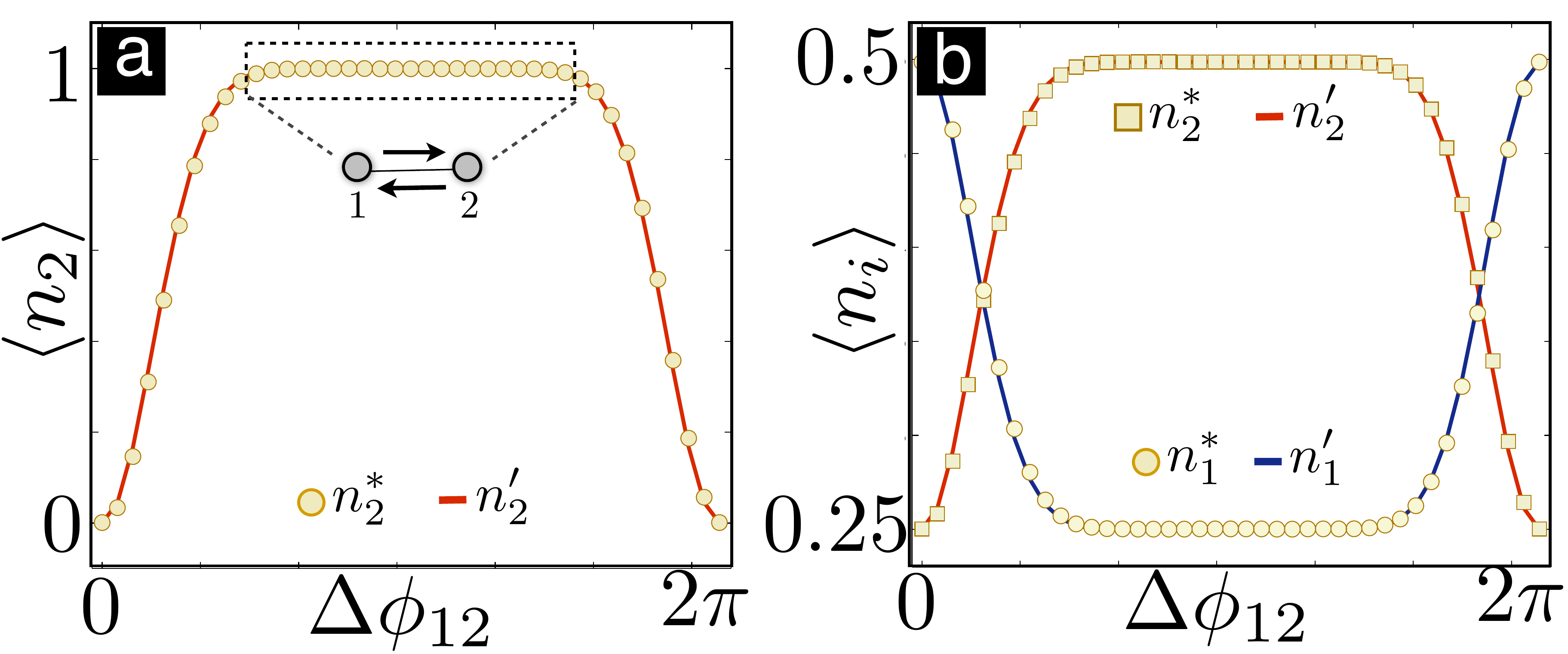}
\caption{ {\bf Phase-dependence and thermal photon-assisted tunneling:} {\bf (a)} Dependence of the PAT on the relative driving phase $\Delta\phi_{12}$. The yellow circles represent the maximum population transfer to site 2, $n_2^*$, for a particle initially located at site 1, $\ket{\Psi_0}=a_1^{\dagger}\ket{\rm vac}$, as obtained from the numerical integration of the full time-dependent Hamiltonian $H_0(t)+H_{\rm t}$. The red line corresponds to the same quantity evaluated for the effective description $H_{\rm eff}$, referred as $n_2'$. Emphasized by the doted frame is the region of optimal assisted tunneling. {\bf (b)} Phase-dependence of the PAT for an initial thermal state  with $\bar{n}_1=0.5,\bar{n}_2=0.25$. We also represent the particle transfer to site 1. }
\label{fig_line_thermal}
\end{figure}

In Fig.~\ref{fig_line_thermal}{\bf (a)}, we study the photon-assisted process for different driving phases $\Delta\phi_{12}\in[0,2\pi]$, considering the same parameters and initial state as before. We represent  the maximum population transferred to site 2 $n_2^*={\rm max}\{\langle n_2(t)\rangle: t\leq t^*\}$, such that $t^*=50\pi/|\mathcal{F}_{1}(\eta_{\rm d},\eta_{\rm d},\pi)|\omega$ corresponds to the optimal PAT with $\Delta\phi_{12}=\pi$. The yellow dots correspond to the numerical integration of the truncated Hamiltonian~\eqref{tight_binding_general} and~\eqref{driving}, whereas the solid line corresponds to the analytical prediction evaluated at the related exchange period $n_2'=\langle n_2(t')\rangle_{\rm eff}$ for the different phases, namely $t'=50\pi/|\mathcal{F}_{1}(\eta_{\rm d},\eta_{\rm d},\Delta\phi_{12})|\omega$. In this figure, we reveal the agreement between both descriptions for any of the driving phases. Note that this agreement will not be compromised by going to larger arrays. As announced previously, we can interpolate between the optimally-assisted and the completely-suppressed tunneling regimes by modifying the driving phase. 

It is also interesting to consider an initial state that does not correspond to a single particle, but rather to a thermal ensemble. We focus now on bosons with independent mean number of particles $\bar{n}_1,\bar{n}_2$. This state corresponds to $\rho_0=\rho_1\otimes\rho_2$ with 
\begin{equation}
\rho_i=\frac{1}{\mathcal{Z}_i}\ee^{-\beta_i\omega_ia_i^{\dagger}a_i},  \hspace{1ex}\mathcal{Z}_i={\rm tr}\big(\ee^{-\beta_i\omega_ia_i^{\dagger}a_i}\big),
\end{equation}
such that the parameters $\beta_i$ are implicitly defined through the mean number of particles following a Bose-Einstein distribution $\bar{n}_i=1/(\ee^{\beta_i\omega_i}-1)$. In this case, we must integrate numerically the Liouville-Von Neumann equation $i{\rm d}\rho(t)/{\rm d}t=[H_0(t)+H_{\rm t},\rho(t)]$, and compare the result to the effective description $\rho_{\rm eff}(t)=\ee^{-i H_{\rm eff}t}\rho_0\ee^{+i H_{\rm eff}t}$. In particular, we consider the same parameters as before, and set $\bar{n}_1=0.5,\bar{n}_2=0.25$ after truncating the particle Hilbert space to $n_{\rm trunc}=7$. Note that due to the thermal effects, the truncation parameter has been increased with respect to the previous simulations. Again, we have checked the convergence of the results by increasing $n_{\rm trunc}$. In Fig.~\ref{fig_line_thermal}{\bf (b)}, we represent the expectation values $n_2^*, n_2'$ introduced above, together with those of site 1, $n_1^*={\rm min}\{\langle n_1(t)\rangle: t\leq t^*\}$, and $n_1'=\langle n_1(t')\rangle_{\rm eff}$ for $t'=50\pi/|\mathcal{F}_{1}(\eta_{\rm d},\eta_{\rm d},\Delta\phi_{12})|\omega$. 
We conclude from the perfect agreement shown in the figure that the scheme also works for thermal states. In fact, in the absence of interactions, the equations of motion do not depend on whether the state is pure or a mixed thermal state. For the latter, there is a background over which the PAT phenomena will occur. The role of the additional interactions, and its interplay with the thermal states,  is an extremely interesting topic that deserves a separate study.

The suitability of the PAT scheme for thermal states will turn out to be essential in Sec.~\ref{pat_ions}, where we discuss a realistic implementation of the quantum simulator of gauge fields with phonons in microtrap arrays. In this case, it means that cooling to the vibrational ground-state is not necessary to observe the non-trivial effects of the PAT.  Another interesting topic is the occurrence of motional heating in the microtraps. In case this motional heating is global, it should not interfere with the PAT effects. However, if this heating has a local nature, it may lead to very interesting effects that mimic the role played by a  reservoir  providing electrons to local regions of a metallic conductor. This effect can be complemented by the controlled engineering of dephasing in the phonon dynamics by introducing noisy potentials in the trap electrodes~\cite{engineered_env}. Hence,  the applicability of the PAT quantum simulator could be widened, and used to study how transport phenomena is affected by an environment that induces decoherence effects such as motional heating or dephasing.

\vspace{1ex}

\subsubsection{ Photon-assisted tunneling around a plaquette}

 Let us  now apply our formalism to a two-dimensional setup where to test the application of PAT for the quantum simulation of synthetic gauge fields. We consider a model of spinless bosons in a square plaquette, and set the parameters of the total Hamiltonian in Eqs.~\eqref{tight_binding_general} and~\eqref{driving} to $J_{{\rm t};12}=10^{-2}\omega$,   $\Delta\omega=0.5\omega$, $\eta_{\rm d}=1$, $r=1$, and $\omega_{\rm d}=\Delta\omega$. Our aim  is to test  the accuracy of the effective description~\eqref{peierls}, trying to find clear-cut evidence of the underlying synthetic gauge field. In order to do so,  we study
a discrete version of the celebrated  Aharonov-Bohm effect~\cite{ab_phase}. 

\begin{figure}
\centering
\includegraphics[width=1\columnwidth]{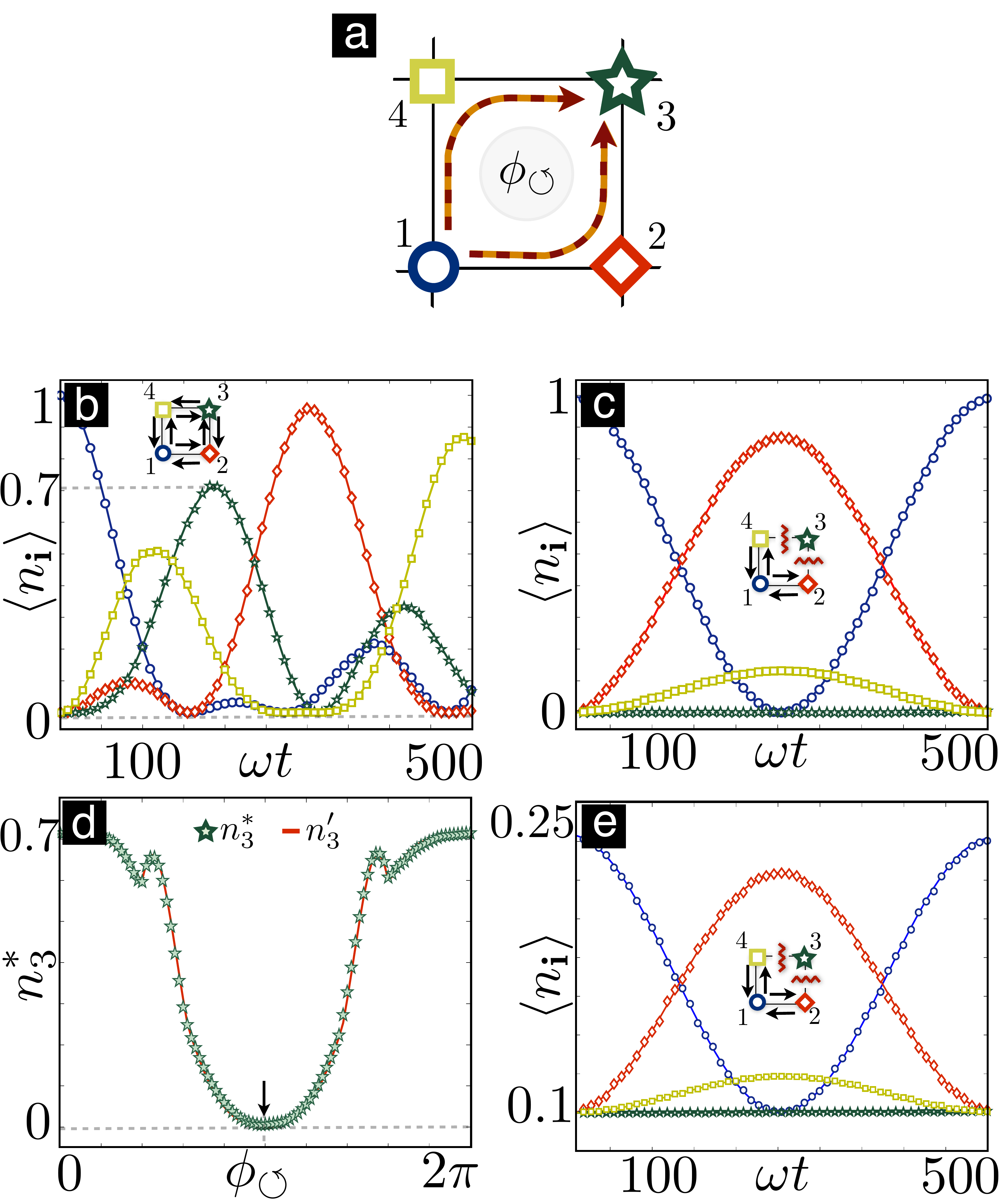}
\caption{ {\bf Photon-assisted tunneling around a plaquette:} {\bf (a)} Schematic representation of the plaquette, highlighting the two possible paths connecting sites 1 and 3. {\bf (b)} PAT for an initial state $\ket{\Psi_0}=a_1^{\dagger}\ket{\rm vac}$, such that the periodic-driving phase is $\phi_1=\pi$, and $\phi_{\circlearrowleft}=r\phi_2=0$ with $r=1$. The numerical expectation values $\langle n_1(t)\rangle$ (blue circles), $\langle n_2(t)\rangle$ (red diamonds), $\langle n_3(t)\rangle$ (green stars), and $\langle n_4(t)\rangle$ (yellow squares) are compared to the effective description in Eq.~\eqref{peierls} (solid lines with same colors).  For vanishing $\phi_{\circlearrowleft}=0$, the particle can tunnel anywhere in the lattice. {\bf (c)} Same as above, but setting  $\phi_{\circlearrowleft}=\pi$, where one clearly observes the Aharonov-Bohm interference $\langle n_3(t)\rangle=0$. {\bf (d)} Maximum population transfer to site 3 as a function of the enclosed phase $\phi_{\circlearrowleft}\in[0,2\pi]$. The yellow dots correspond to the exact numerical integration $n_3^*$, and the solid red line to the effective description $n_3'$. A black arrow marks the perfect Aharonov-Bohm interference.  {\bf (e)} Aharonov-Bohm interference for an initial thermal state $\bar{n}_1=0.25$, and $\bar{n}_2=\bar{n}_3=\bar{n}_4=0.1$        }
\label{fig_square}
\end{figure}

Let us briefly recall the underlying interference effect [Fig.~\ref{fig_square}{\bf (a)}]. A particle initially localized at site 1 can tunnel to site 3 following two different paths, namely $\gamma_1:1\rightarrow 2\rightarrow 3$, or $\gamma_2:1\rightarrow 4\rightarrow 3$. Due to the  gauge field, each path leads to a different phase ${\ket{\Psi_{\rm f}}}\propto \ee^{i e^*\int_{\gamma_1}{\rm d}{\bf r}\cdot {\bf A}_{\rm s}}a_3^{\dagger}\ket{{\rm vac}}+\ee^{ie^*\int_{\gamma_2}{\rm d}{\bf r}\cdot {\bf A}_{\rm s}}a_3^{\dagger}\ket{\rm vac}$, such that  the probability of performing such trajectory is $P_3\propto 2+2\cos\big(e^*\int_{\gamma_1}{\rm d}{\bf r}\cdot {\bf A}_{\rm s}-e^*\int_{\gamma_2}{\rm d}{\bf r}\cdot {\bf A}_{\rm s}\big)=2(1+\cos \phi_{\circlearrowleft})$, where $\phi_{\circlearrowleft}=e^*\oint_{\circlearrowleft}{\rm d}{\bf r}\cdot {\bf A}_{\rm s}$ was defined in Eq.~\eqref{ab_phase}. From this expression, one finds a perfect destructive interference that forbids the particle to tunnel to site $3$ when the  enclosed phase  $\phi_{\circlearrowleft}=\pi$. This Aharonov-Bohm interference will serve us as a  testbed for the validity of the gauge-field analogy~\eqref{peierls}.

We now compare the dynamics obtained from the numerical integration of the driven Hamiltonian~\eqref{tight_binding_general} and~\eqref{driving}, to the effective description in Eq.~\eqref{peierls}. In Figs.~\ref{fig_square}{\bf (b)-(c)}, we consider  $\ket{\Psi_0}=a_1^{\dagger}\ket{\rm vac}$, and study the propagation of the particle along the square plaquette for different driving phases. In Fig.~\ref{fig_square}{\bf (b)}, we observe that for the phases $\phi_1=\pi,\phi_2=0$, the particle is allowed to tunnel to every lattice site. This is consistent with the fact that for $\phi_{\circlearrowleft}=\phi_2=0$, there is no interference. Conversely, we set $\phi_1=\pi,\phi_2=\pi$ in Fig.~\ref{fig_square}{\bf (c)}, where one readily observes that the tunneling to site $3$ is forbidden by the aforementioned Aharonov-Bohm interference. We note that for the trapped-ion case (see the sections below), the diagonal path going from sites $1\to3$ and $2\to4$ must also be accounted since the tunneling amplitude need not be  small. In~\cite{gauge}, we discussed how to cancel it by playing with the phases, so that the analogy with the Aharonov-Bohm effect is valid. In Fig.~\ref{fig_square}{\bf (e)}, we corroborate that the effect also holds for thermal states. Finally, we check in Fig.~\ref{fig_square}{\bf (d)} that the perfect destructive interference only occurs for the so-called $\pi$-flux phase. In this figure, we represent the maximal transferred population to site 3, $n_3^*={\rm max}\{\langle n_3(t),0<t<t^*\}$, where $t^*=100\pi/|\mathcal{F}_1(\eta_{\rm d},\eta_{\rm d},\pi)|\omega$, and show that the perfect interference only occurs when the enclosed phase $\phi_{\circlearrowleft}=\pi$. This allows us to rule out other possible sources of interference, and conclude that it is only due to the synthetic gauge field.

Let us close this subsection by highlighting the accuracy of our analytical treatment, which has been confronted to a numerical survey of different setups with a wide range of parameters.  The rest of this manuscript will be devoted to analyze the experimental setups in the field of quantum optics,  ions in microtrap arrays, where the necessary ingredients for this PAT are in reach with state-of-the-art technologies. Besides, we shall also introduce novel  methods to to exploit additional aspects of  PAT phenomenology.

\section{Photon-assisted-tunneling toolbox for  trapped ions}
\label{pat_ions}
In this section, we apply the above scheme to arrays of  micro-fabricated ion traps, and show that one can achieve PAT of the vibrational excitations between distant microtraps. In order to set the notation, we start by showing in Sec.~\ref{phonon_tunneling} that the vibrations of an ensemble  of ions in a microtrap array can be described in terms of tunneling phonons~\cite{hubbard_porras}. Then, we describe in Sec.~\ref{driving_ions} how the gradient and periodic driving of the trapping frequencies can be achieved using the tools of state-of-the-art experiments in ion traps (see the reviews~\cite{ion_trap_reviews}). In particular, the gradient can be implemented by the  local control of DC-voltages applied to the trap electrodes, whereas the periodic driving stems from an optical dipole force. We derive a set of constraints that this dipole force must fulfill, and discuss how these requirements can be met for realistic experimental parameters. In Sec.~\ref{synthetic_int}, we incorporate phonon-phonon interactions in the phonon-based quantum simulator of bosonic particles exposed to synthetic gauge fields. Finally,  in Secs.~\ref{non_abelian}, \ref{disorder}, and \ref{flux_lattices}, we describe how to extend the  PAT scheme beyond the applications discussed so far, reaching more involved models where the spin  of the ion can be exploited as  an additional tool.

\subsection{Tunneling of phonons in microtrap arrays}
\label{phonon_tunneling}
Let us consider an ensemble of $N$  ions with charge $e$ and mass $m$, labelled by  integer numbers ${\bf i}=(i_x,i_y)$. Each ion is subjected to the electric potential energy created by a planar micro-fabricated trap $V_{\rm t}$, and the Coulomb energy potential $V_{\rm c}$ causing the repulsion between the remaining ions, such that   the total Hamiltonian is 
\begin{equation}
H=H_0+H_{\rm c}=\sum_{{\bf i}}\left(\frac{{\bf p}_{{\bf i}}^2}{2m}+V_{\rm t}({\bf r}_{\bf i})\right)+\frac{e^2}{2}\sum_{{\bf i}\neq{\bf j}}\frac{1}{|{\bf r}_{\bf i}-{\bf r}_{\bf j}|}.
\end{equation}
Here, the trap is designed in such a way that the ion equilibrium positions, which are given by $\boldsymbol{\nabla} (V_{\rm t}+V_{\rm c})|_{\bf r_{\bf i}^0}={\bf 0}$, form  a regular lattice ${\bf r}_{\bf i}^0=i_1d_1{\bf e}_1+i_2d_2{\bf e}_2$, where ${\bf e}_{1},{\bf e}_{2}$  are the unit vectors, and $d_{1},d_{2}$ the lattice spacings. At sufficiently small temperatures, the excursions of the ions from these equilibrium positions ${\bf r}_{\bf i}={\bf r}_{\bf i}^0+\delta{\bf r}_{\bf i}$ are small with respect to the lattice spacing. Thus, for our purpose, we are allowed to consider the above Hamiltonian up to second order only~\cite{feynman_lectures,james}. This  approximation yields a system of coupled harmonic oscillators
\begin{equation}
\label{coupled_oscillators}
H=\sum_{\bf i}\frac{{\bf p}_{\bf i}^2}{2m}+\sum_{{\bf i}\neq{\bf j} }\sum_{\alpha,\gamma}\big(\mathcal{V}_{{\rm t};{\bf i} {\bf j}}^{\alpha\gamma}+\mathcal{V}_{{\rm c};{\bf i} {\bf j}}^{\alpha\gamma}\big)\delta r_{\alpha,{\bf i}}\delta r_{\gamma,{\bf j}},
\end{equation}
with the following couplings  
\begin{equation}
\label{coulomb}
\begin{split}
\mathcal{V}_{{\rm t};{\bf i} {\bf j}}^{\alpha\gamma}&=\left.\frac{1}{2}\frac{\partial^2 V_{\rm t}}{\partial r_{\alpha,{\bf i}}\partial r_{\gamma,{\bf j}}}\right|_{\{{\bf r}_{\bf l}^0\}}=\frac{m}{2}\omega_{\alpha,{\bf i}}^2\delta_{{\bf i}{\bf j}}\delta_{\alpha\gamma},\\
\mathcal{V}_{{\rm c};{\bf i} {\bf j}}^{\alpha\gamma}&=\left.\frac{1}{2}\frac{\partial^2 V_{\rm c}}{\partial r_{\alpha,{\bf i}}\partial r_{\gamma,{\bf j}}}\right|_{\{{\bf r}_{\bf l}^0\}}=\\
&=\frac{e^2}{2}\sum_{{\bf l}\neq{\bf i}}(\delta_{{\bf l}{\bf j}}-\delta_{{\bf i}{\bf j}})\left(\frac{\delta_{\alpha\gamma}}{|{{\bf r}^0_{\bf i-l}}|^3}-\frac{3 \left( {\bf r}^0_{\bf i-l}\right)_{\alpha}\left({\bf r}^0_{\bf i-l}\right)_{\gamma}}{|{{\bf r}^0_{\bf i-l}}|^5}\right),
\end{split}
\end{equation}
where  ${\bf r}^0_{\bf i-l}={\bf r}^0_{\bf i}-{\bf r}^0_{\bf l}$, $\delta_{{\bf i}{\bf l}}$ stands for the Kronecker delta, and $\alpha,\gamma=x,y,z$ refer to the main axes of the trapping potential. Note that the   micro-fabricated trap gives rise to a confinement that can be considered to be harmonic for the motional amplitudes considered here, and is  therefore characterized by  the frequencies $\omega_{\alpha,{\bf i}}$ that may depend on the axis and the  ion position  within the lattice. The Hamiltonian~\eqref{coupled_oscillators} can be expressed in the basis of local quantized vibrations
\begin{equation}
\begin{split}
\delta r_{\alpha,{\bf i}}=&\sqrt{\frac{1}{2m\omega_{\alpha,{\bf i}}}}\left(b_{\alpha,{\bf i}}^{\dagger}+b_{\alpha,{\bf i}}^{\phantom{\dagger}}\right),\\
p_{\alpha,{\bf i}}=i&\sqrt{\frac{\phantom{2}m\omega_{\alpha,{\bf i}}}{2}}\left(b_{\alpha,{\bf i}}^{\dagger}-b_{\alpha,{\bf i}}^{\phantom{\dagger}}\right),
\end{split}
\end{equation}
where $b_{\alpha,{\bf i}}^{\dagger}, b_{\alpha,{\bf i}}^{\phantom{\dagger}}$ are the bosonic creation-annihilation operators, and we set $\hbar=1$. In this local basis, the vibrational Hamiltonian 
becomes $H=H_0+H_{\rm c}$,
\begin{equation}
\label{tight_binding}
H=\sum_{\alpha,{\bf i}}\omega_{\alpha,{\bf i}}b_{\alpha,{\bf i}}^{\dagger}b_{\alpha,{\bf i}}^{\phantom{\dagger}}+\sum_{{\bf i}{\bf j}}\sum_{\alpha,\gamma}J_{{\rm c};{\bf i j}}^{\alpha\gamma}\big(b_{\alpha,{\bf i}}^{\dagger}+b^{\phantom{\dagger}}_{\alpha,{\bf i}}\big)\big(b_{\gamma,{\bf j}}^{\dagger}+b^{\phantom{\dagger}}_{\gamma,{\bf j}}\big),
\end{equation}
where the Coulomb couplings $ J_{{\rm c};{\bf ij}}^{\alpha\gamma}=\mathcal{V}^{\alpha\gamma}_{{\rm c};{\bf i}{\bf j}}/(2m\sqrt{\omega_{\alpha,{\bf i}}\omega_{\gamma,{\bf j}}})
$  describe the exchange of vibrational excitations between different ions ${\bf i},\alpha\leftrightarrow{\bf j},\gamma$, and yield the aforementioned collective phonons once diagonalized. Let us express $H_{\rm c}$ in the interaction picture with respect to $H_0$, namely $H_{\rm c}'=\ee^{iH_0t}H_{\rm c}\ee^{-iH_0t}$, 
\begin{equation}
\label{coulomb_mixing}
H_{\rm c}'=\hspace{-0.5ex}\sum_{{\bf i}{\bf j}\alpha\gamma}\hspace{-0.5ex}J_{{\rm c};{\bf i j}}^{\alpha\gamma}\big(b_{\alpha,{\bf i}}^{\dagger}b^{\phantom{\dagger}}_{\gamma,{\bf j}}\ee^{i(\omega_{\alpha,{\bf i}}-\omega_{\gamma,{\bf j}})t}+b_{\alpha,{\bf i}}^{\dagger}b_{\gamma,{\bf j}}^{\dagger}\ee^{i(\omega_{\alpha,{\bf i}}+\omega_{\gamma,{\bf j}})t}\big)\hspace{-0.5ex}+\text{H.c.}
\end{equation}
In this work, we consider that the microtraps fulfill 
\begin{equation}
\textstyle{\alpha\neq\gamma},\hspace{0.5ex}J_{{\rm c};{\bf i j}}^{\alpha\gamma}\ll |\omega_{\alpha,{\bf i}}-\omega_{\gamma,{\bf j}}|,\hspace{2ex}\alpha=\gamma,\hspace{0.5ex}J_{{\rm c};{\bf i j}}^{\alpha\alpha}\ll |\omega_{\alpha,{\bf i}}+\omega_{\alpha,{\bf j}}|,
\end{equation} 
which allow us to neglect the rapidly oscillating terms of the above Hamiltonian using a standard rotating-wave approximation (RWA). The first condition allows us to consider  independent vibrations along each of the trapping axes, whereas the second one allows us to neglect the terms that do not conserve the number of vibrational excitations. Accordingly,
\begin{equation}
\label{tight_binding_bis}
H=H_0+H_{\rm c}=\sum_{{\bf i},\alpha}\omega_{{\alpha},{\bf i}}'b_{{\alpha},{\bf i}}^{\dagger}b_{{\alpha},{\bf i}}^{\phantom{\dagger}}+\sum_{\alpha}\sum_{{\bf i}>{\bf j}}\big(J_{{\rm c};{\bf i j}}^{{\alpha}}b_{{\alpha},{\bf i}}^{\dagger}b_{{\alpha},{\bf j}}^{\phantom{\dagger}}+\text{H.c.}\big),
\end{equation}
where the trapping frequency of each ion is slightly modified by the electrostatic interaction with its neighboring ions $\omega_{{\alpha},{\bf i}}'=(\omega_{{\alpha},{\bf i}}^2+\mathcal{V}^{{\alpha}{\alpha}}_{{\rm c};\bf ii}/2m)^{1/2}$, and we have defined $J_{{\rm c};{\bf i j}}^{{\alpha}}=2J_{{\rm c};{\bf i j}}^{{\alpha}{\alpha}}$.
 This Hamiltonian can be interpreted as a tight-binding model for the local phonons, which tunnel between distant microtraps according to  the dipolar tunneling strengths $J_{{\rm c};{\bf i j}}^{{\alpha}}$. By a direct comparison to the general tight-binding model in Eq.~\eqref{tight_binding_general}, we can identify the index $\sigma$ with the vibrational axis, the tunneling strength $J_{{\rm t};{\bf i}{\bf j}}^{\sigma}$ with the aforementioned dipolar Coulomb couplings, and the on-site energies $\omega_{\sigma,{\bf i}}$ with the microtrap trapping frequencies. In the following subsection, we  describe how to obtain the gradient and the periodic driving using state-of-the-art tools in trapped-ion experiments.

\subsection{Gradient and periodic driving of the trapping frequencies}
\label{driving_ions}
In  microtrap arrays, it is possible to design the individual trapping frequencies $\{\omega_{\alpha,{\bf i}}\}$ by the  control  of DC-voltages applied to local micro-fabricated electrodes. In fact, the experiments~\cite{inhibitted_hopping_exp} have made explicit use of this property in a conventionally segmented linear rf-trap, in order to switch on/off the phonon tunneling by tuning the trapping frequencies of two neighboring  traps in/out of common resonance~\cite{microtraps}.  We chose the trapping frequencies  distributed according to a gradient $\omega_{{\alpha},{\bf i}}=\omega_{{\alpha}}+\Delta\omega_{{\alpha}}i_1$. Considering mutual trap and mutual ion distances $>10\mu$m due to current constraints in fabrication. The correction to the trapping frequency due to the electrostatic interaction can be neglected, and we get the desired gradient of the on-site energies $\omega_{{\alpha},{\bf i}}'\approx\omega_{{\alpha},{\bf i}}=\omega_{{\alpha}}+\Delta\omega_{{\alpha}}i_1$ which must be incorporated to the tight-binding Hamiltonian~\eqref{tight_binding_bis}.

\begin{figure}
\centering
\includegraphics[width=1\columnwidth]{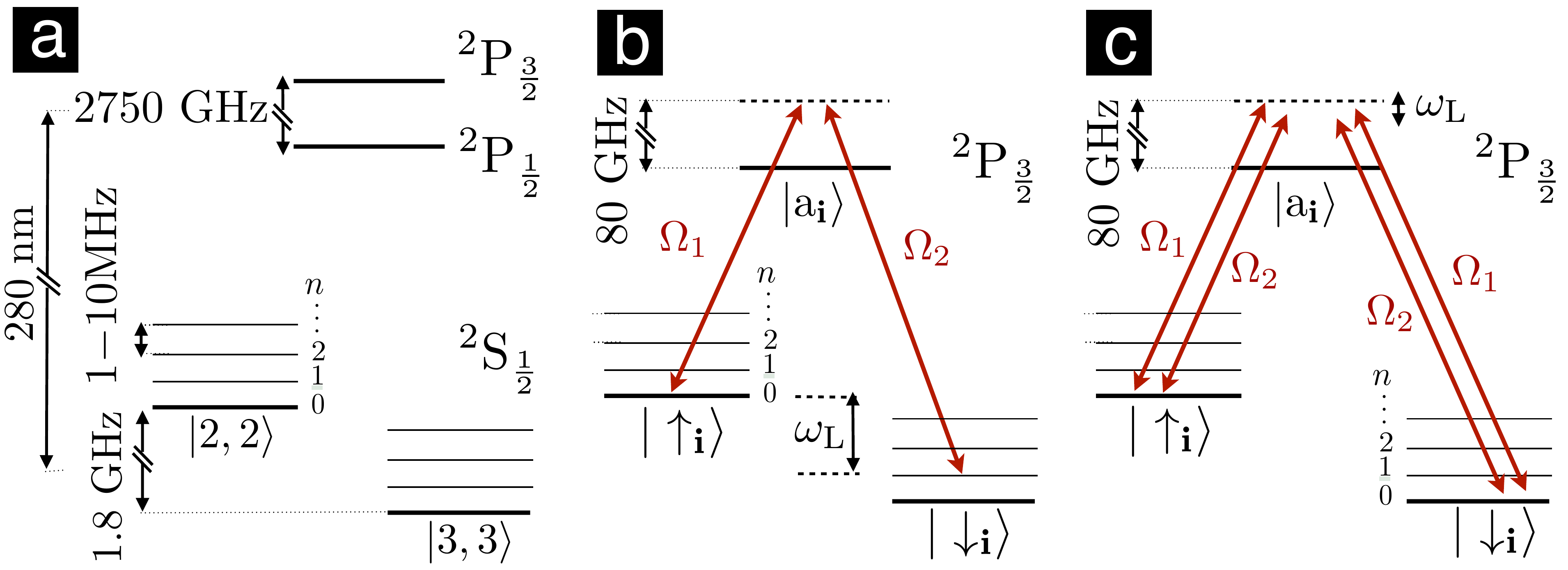}
\caption{ {\bf Energy-level scheme and laser-beam arrangement:} {\bf (a)} Reduced diagram for the  hyperfine energy levels $\ket{F,M}$ of $^{25}{\rm Mg}^+$. The ground-state manifold $^2{\rm S}_{1/2}$ is split into the hyperfine levels $F=2,3$, such that an additional magnetic field allows us to isolate two magnetic sublevels  $\ket{\uparrow_{\bf i}}=\ket{2,2}_{\bf i}, \ket{\downarrow_{\bf i}}=\ket{3,3}_{\bf i}$ with an energy difference of $\omega_0/2\pi\approx 1.8$ GHz. Note also that these levels are dressed by a vibrational ladder with equidistant spacing $\omega_{\alpha}/2\pi\approx 1-10$MHz. The excited manifold $^2{\rm P}_{3/2}$ contains an auxiliary level to implement a stimulated  Raman scheme with two laser beams and a beatnote $\omega_{\rm L}=\omega_1-\omega_2$. {\bf (b)} Raman scheme for two laser beams with Rabi frequencies $\Omega_1,\Omega_2$ in the so-called red-sideband regime $\omega_{\rm L}\approx\omega_0-\omega_{\alpha}$. In this regime, excitations are exchanged between the electronic and vibrational states of the ions. {\bf (c)} Raman scheme for the periodic modulation of the trapping frequencies $\omega_{\rm L}\ll\omega_{\alpha}$. In this limit, only negligible coupling between the internal or vibrational states take place during the transition, but rather a periodic modulation of the trapping frequencies is realized.}
\label{fig_laser}
\end{figure}

We now turn to the periodic driving of the trapping frequencies, which shall be implemented by an optical dipole force~\cite{ion_trap_reviews}. We consider a pair of laser beams with frequencies $\omega_1,\omega_2$, and wavevectors ${\bf k}_1,{\bf k}_2$, which couple to the electronic states $\ket{\rm \downarrow_{\bf i}},\ket{\rm \uparrow_{\bf i}}, \ket{\rm a_{\bf i}}$ (see the energy-level scheme of Fig.~\ref{fig_laser} for the particular case of $^{25}$Mg$^+$ ions). Encoding the spin degree of freedom into a pair of electronic ground states by exploiting their hyperfine or Zeeman splitting, the energy difference $\omega_0/2\pi$  between the  states $\ket{\rm \downarrow_{\bf i}},\ket{\rm \uparrow_{\bf i}}$ lies in the microwave range $1$ GHz, whereas  the  energy splitting $\omega_{\rm a}/2\pi$ to an auxiliary state $\ket{\rm a_{\bf i}}$ is much larger $\omega_{\rm a}\gg\omega_0$, and typically lies in the optical domain $100-1000$ THz.  By tuning the beatnote of the phase-locked laser beams to a particular value, it is possible to choose out of a  variety of couplings. For instance, when $\omega_1-\omega_2\approx \omega_0$, such that $\omega_1,\omega_2\ll\omega_{\rm a}$, one speaks about a stimulated two-photon Raman transition between the ground and excited states [Fig.~\ref{fig_laser}{\bf (b)}]. Conversely, when $\omega_1-\omega_2\approx \omega_{\alpha}\ll\omega_0$, one obtains a running-wave realization of the so-called spin-dependent dipole forces. In this work, we are interested in yet a different regime  $\omega_1-\omega_2\ll \omega_{{\alpha}}$ [Fig.~\ref{fig_laser}{\bf (c)}], where we get the desired periodic driving by the laser dipole force. We consider  the ion-laser interaction in the dipolar approximation
\begin{equation}
\label{laser_ion}
\begin{split}
H_{\rm L}=&\sum_{l=1,2}\sum_{{\bf i}}\half\Omega^{(l)}_{\rm a\uparrow}\ket{{\rm a}_{\bf i}}\bra{{\uparrow}_{\bf i}}\ee^{-i({\bf k}_l{\bf r_{\bf i}}-\omega_lt)}+\text{H.c.}\\
+&\sum_{l=1,2}\sum_{{\bf i}}\half\Omega^{(l)}_{\rm a\downarrow}\ket{{\rm a}_{\bf i}}\bra{{\downarrow}_{\bf i}}\ee^{-i({\bf k}_l{\bf r_{\bf i}}-\omega_lt)}+\text{H.c.},
\end{split}
\end{equation}
where $\Omega^{(l)}_{\rm a\uparrow},\Omega^{(l)}_{\rm a\downarrow}$ are the Rabi frequencies for each of the transitions to the auxiliary level induced by each of the laser beams. Since we are assuming that the laser frequencies are far off-resonant with respect to the auxiliary state, it is possible to perform an adiabatic elimination of this state to simplify the dynamics between $\ket{\uparrow_{\bf i}},\ket{\downarrow_{\bf i}}$. Furthermore, for sufficiently small Rabi frequencies and considering that $\omega_1-\omega_2\ll\omega_{0}$, it is possible to obtain 
\begin{equation}
\label{raman}
\begin{split}
H_{\rm L}=&-\sum_{\bf i}\left[\textstyle{\frac{\big(\Omega^{(1)}_{\rm a\downarrow}\big)^*\Omega^{(2)}_{\rm a\downarrow}}{4\Delta}\ee^{i(\Delta{\bf k}\cdot{\bf r}_{\bf i}-\omega_{\rm L}t)}+\text{H.c.}}\right]\ket{{\downarrow}_{\bf i}}\bra{{\downarrow}_{\bf i}}\\
&-\sum_{\bf i}\left[\textstyle{\frac{\big(\Omega^{(1)}_{\rm a\uparrow}\big)^*\Omega^{(2)}_{\rm a\uparrow}}{4\Delta}\ee^{i(\Delta{\bf k}\cdot{\bf r}_{\bf i}-\omega_{\rm L}t)}+\text{H.c.}}\right]\ket{{\uparrow}_{\bf i}}\bra{{\uparrow}_{\bf i}},
\end{split}
\end{equation}  
where we have defined the beatnote $\omega_{\rm L}=\omega_1-\omega_2$,  $\Delta {\bf k}={\bf k}_1-{\bf k}_2$, and $\Delta=\omega_{\rm a}-\omega_0-\omega_1$ such that $|\Delta|\gg\omega_0$. Note that we have neglected the AC-Stark shifts arising from each single laser beam, since they do not play any role for the periodic driving of the trapping frequencies. By a proper choice of the laser intensities, detunings, and polarizations, one finds a regime where the two-photon Rabi frequency is $\Omega_{\rm L}= -\big(\Omega^{(1)}_{\rm a\downarrow}\big)^*\Omega^{(2)}_{\rm a\downarrow}/2\Delta=-\big(\Omega^{(1)}_{\rm a\uparrow}\big)^*\Omega^{(2)}_{\rm a\uparrow}/2\Delta$, and thus the laser-ion interaction becomes
\begin{equation}
\label{stark_shift}
H_{\rm L}=\half\Omega_{\rm L}\sum_{\bf i}\ee^{i(\Delta {\bf k}\cdot{\bf r}_{\bf i}-\omega_{\rm L}t)}+\text{H.c.}
\end{equation}
This expression corresponds to a Stark shift that acts equally on both electronic states, and is caused by  the crossed laser beams of different frequencies.

To obtain the desired periodic driving from the ion-laser Hamiltonian~\eqref{stark_shift}, we express the ion position in terms of the local phonon operators ${\bf r}_{\bf i}={\bf r}_{\bf i}^0+\sum_{\alpha}{\bf e}_{\alpha}(b_{\alpha,{\bf i}}^{\dagger}+b_{\alpha,{\bf i}}^{\phantom{\dagger}})/\sqrt{2m\omega_{\alpha}}$. By introducing the so-called Lamb-Dicke parameter $\eta_{\alpha}={\bf e}_{\alpha}\cdot\Delta {\bf k}/\sqrt{2m\omega_{\alpha}}\ll 1$, we can Taylor expand the Hamiltonian $H_{\rm L}\approx H_{\rm L}^{(0)}+H_{\rm L}^{(1)}+H_{\rm L}^{(2)}$, whereby
\begin{equation}
\label{taylor}
\begin{split}
H_{\rm L}^{(0)}&=\frac{\Omega_{\rm L}}{2}\sum_{\bf i}\ee^{i\Delta{\bf k}\cdot {\bf r}_{\bf i}^0-i\omega_{\rm L}t}+\text{H.c.} \\
H_{\rm L}^{(1)}&=\frac{i\Omega_{\rm L}}{2}\sum_{{\bf i}\alpha}\ee^{i\Delta{\bf k}\cdot {\bf r}_{\bf i}^0-i\omega_{\rm L}t}\eta_{\alpha}(b_{\alpha,{\bf i}}^{\phantom{\dagger}}+b_{\alpha,{\bf i}}^{\dagger})+\text{H.c.}\\
H_{\rm L}^{(2)}&=\frac{-\Omega_{\rm L}}{4}\sum_{{\bf i}\alpha\gamma}\ee^{i\Delta{\bf k}\cdot {\bf r}_{\bf i}^0-i\omega_{\rm L}t}\eta_{\alpha}\eta_{\gamma}(b_{\alpha,{\bf i}}^{\phantom{\dagger}}+b_{\alpha,{\bf i}}^{\dagger}) (b_{\gamma,{\bf i}}^{\phantom{\dagger}}+b_{\gamma,{\bf i}}^{\dagger})\\
&+\text{H.c.}\\
\end{split}
\end{equation}   
Note that the first term corresponds to an irrelevant $c$-number that does not modify the phonon dynamics. In order to find the relevant contribution of the two remaining terms $H_{\rm L}^{(1)},H_{\rm L}^{(2)}$, we switch to the interaction picture with respect to $H_0=\sum_{\alpha,{\bf i}}\omega_{\alpha,{\bf i}}b^{\dagger}_{\alpha,{\bf i}}b_{\alpha,{\bf i}}^{\phantom{\dagger}}$. If we consider that both, the frequency gradient and the laser frequency, fulfill $\Delta\omega_{\alpha},\omega_{\rm L}\ll \omega_{\alpha},|\omega_{\alpha}-\omega_{\gamma}|_{\alpha\neq\gamma}$, a rotating wave approximation for $\eta_{\alpha}\Omega_{\rm L}\ll \omega_{\alpha}$ allow us to neglect the rapidly-oscillating terms and leads to the desired periodic driving of the trapping frequencies
\begin{equation}
\label{periodic_driving}
H_{\rm L}\approx-\Omega_{\rm L}\sum_{\alpha,{\bf i}} \eta_{\alpha}^2\cos(\Delta{\bf k}\cdot{\bf r}_{\bf i}^0-\omega_{\rm L}t)b^{\dagger}_{\alpha,{\bf i}}b_{\alpha,{\bf i}}^{\phantom{\dagger}}.
\end{equation}
By direct comparison with Eq.~\eqref{driving}, we find the following, individually tunable, mapping between  the parameters of the periodic driving and the laser beams:  $\omega_{\rm d}=\omega_{\rm L}$,  $\omega_{\rm d}\eta_{\rm d}=-\Omega_{\rm L}\eta_{\alpha}^2$, and $\phi_{\bf i}=-\Delta{\bf k}\cdot {\bf r}_{{\bf i}}^0$. Let us also note that in order to neglect the higher order terms that rotate with the laser frequency $\omega_{\rm L}$, we impose that $\Omega_{\rm L}\eta^4_{\alpha}\ll \omega_{\rm L}$. It is important to remark that even if we are considering small  Lamb-Dicke parameters $\eta_{\alpha}\ll 1$, the periodic driving $|\eta_{\rm d}|=\Omega_{\rm L}\eta_{\alpha}^2/\omega_{\rm L}$ needs not to be small by selecting the appropriate laser and Rabi frequencies, so that the non-perturbative character of the PAT can be exploited. 

\begin{table}
   \caption{Typical parameters for ion microtrap arrays.}
 \begin{ruledtabular}
  \begin{tabular}{cccccc}
$\omega_{\alpha}/2\pi$& $\Delta\omega_{\alpha}/2\pi$ & $\omega_{\rm L}/2\pi$&$J_{{\rm c};12}^{\alpha}/2\pi$ & $\Omega_{\rm L}/2\pi$ &$\eta_{\alpha}$ \\
  \hline
1-10 MHz & 50-500 kHz &  50-500 kHz& 1-10 kHz &0.1-1 MHz & 0.1-0.4\\ 
  \end{tabular}
   \end{ruledtabular} 
   \label{ti_parameters}
   \end{table}

The remaining task to demonstrate that the scheme of PAT  works for phonons in microtrap arrays is to consider the typical values for the experimental parameters, and discuss whether they satisfy the constraints imposed during the above derivation. The orders of magnitude of all the relevant parameters are listed in Table~\ref{ti_parameters}, which satisfy the different constraints  made along this derivation, namely 
\begin{equation}
\label{constraints}
 \eta_{\alpha}\Omega_{\rm L}\ll \omega_{\alpha}, \hspace{1ex} \Delta\omega_{\alpha},\omega_{\rm L}\ll \omega_{\alpha},|\omega_{\alpha}-\omega_{\gamma}|_{\alpha\neq\gamma}.
\end{equation} 
For the last inequality, it suffices to focus on the phonon modes transverse to the microtrap plane, ${\alpha}=z$, such that $|\omega_{z}-\omega_{\gamma}|_{\gamma\neq z}/2\pi\approx 1$MHz. If the in-plane vibrational modes are to be used, the anisotropy of the trapping frequencies $\omega_x\neq\omega_y$ should be considered in the microtrap design.   Finally, table~\ref{ti_parameters} allows us to estimate the parameters of the model Hamiltonian, and reveal whether the constraints for the PAT in Eqs.~\eqref{gradient_constraint} and~\eqref{driving_constraint} are fulfilled. We find that  $J_{\rm c;12}^{\alpha}/2\pi\approx$ 1-10kHz$\ll\Delta\omega_{\alpha}/2\pi\approx$ 50-500kHz, which thus satisfies the constraint in Eq.~\eqref{gradient_constraint}. Besides, we find that  $\phi_{\bf i}=\phi_1i_1+\phi_2 i_2$, where $\phi_{\alpha}=-(\Delta {\bf k}\cdot {\bf e}_{\alpha})d_{\alpha}$, and thus the constraint~\eqref{driving_constraint} is also satisfied. Therefore, we can conclude that the required ingredients for the PAT introduced in Sec.~\ref{pat_general} can already be met with the current technology of microtrap arrays if the tunneling rates outrun the decoherence rates. Note  that  if motional heating turns out to degrade the results severely, we could mitigate this problem by the using a cryogenic setup.

Let us finally comment on two additional constraints different from~\eqref{gradient_constraint} and~\eqref{driving_constraint}, which are particular to the trapped-ion setup. Both the gradient and the the periodic driving should not modify the stability of the ion crystal, and thus must be smaller than the trapping frequencies
\begin{equation}
\label{stability_constraint}
\Delta\omega_{\alpha},\eta_{\rm d}\omega_{\rm d}\ll\omega_{\alpha}.
\end{equation}
This is also fulfilled for the parameters in Table~\ref{ti_parameters}. Let us note that this condition sets a limit to the scalability of our proposal. For gradients $\Delta\omega_{\alpha}/2\pi\approx$ 50kHz, and trapping frequencies $\omega_{\alpha}/2\pi\approx$1MHz, an array of 10$\times$10 microtraps is  still  consistent with the  maximum attainable trapping frequency. We note  that the gradient can be reduced further (10kHz) without compromising the efficiency of the PAT scheme, and leading to larger systems with $N=$2500 microtraps.  
Finally, we would like to remark that the scheme could be scaled even further by considering a local gradient that only affects a few sites, and repeats periodically along one axes of the microtrap array. In order to assist the tunneling between the sites where the gradient is changed, while maintaining a homogenous synthetic flux, an additional periodic driving with a suitable frequency and phase must also be introduced. 

\subsection{Synthetic gauge fields and phonon-phonon interactions}
\label{synthetic_int}
The discussion of the PAT of phonons has focused on the quadratic  Hamiltonians corresponding to the Coulomb-induced tunneling~\eqref{tight_binding_bis}, and the periodic driving of the trap frequencies~\eqref{periodic_driving}.  However, as discussed in Sec.~\ref{pat_general},  non-linearities corresponding to the on-site interactions in Eq.~\eqref{onsite_interactions} can be incorporated without modifying the assisted-tunneling scheme.  In fact, this broadens the applicability of our many-body quantum simulator, since it also  targets models of strongly-interacting particles. To obtain such phonon-phonon interactions~\cite{hubbard_porras}, strong non-linearities in the trapping potential are required. Remarkably, one can exploit a different realization of the above dipole forces~\eqref{stark_shift} in order to give rise to such non-linearities. We denote the parameters of these new dipole forces by a wiggled bar in order to differentiate them form the dipole forces leading to the periodic driving.

We consider that two additional laser beams  leading to a dipole force have the same frequency, in a way that the beatnote $\tilde{\omega}_{\rm L}=0$, and Eq.~\eqref{stark_shift} corresponds to a standing wave. In analogy to the previous expansion~\eqref{taylor}, we consider that $\tilde{\Omega}_{\rm L}\tilde{\eta}_{\alpha}\ll \omega_{\alpha}$ so that all the terms that do not conserve the number of phonons can be neglected. However, since $\tilde{\omega}_{\rm L}=0$,  the quadratic terms now correspond to a small shift of the trapping frequencies which has no effect since $\tilde{\Omega}_{\rm L}\tilde{\eta}_{\alpha}^2\ll \Delta\omega_{\sigma}\ll \omega_{\alpha}$. The most relevant contribution is due to the quartic terms 
\begin{equation}
\label{onsite_interactions_bis}
\tilde{H}_{\rm L}\hspace{-0.5ex}\approx \hspace{-0.5ex}\sum_{\ii,\alpha,\gamma}\tilde{U}_{{\bf i},\alpha\gamma}b_{\alpha,\ii}^{\dagger}b_{\gamma,\ii}^{\dagger}b_{\gamma,\ii}^{\phantom{\dagger}}b_{\alpha,\ii}^{\phantom{\dagger}},\hspace{0.5ex}\tilde{U}_{{\bf i},\alpha\gamma}=\half\tilde{\Omega}_{\rm L}\tilde{\eta}_{\alpha}^2\tilde{\eta}_{\gamma}^2\cos(\tilde{\Delta{\bf k}}\cdot{\bf r}_{\ii}^0)
\end{equation}
which corresponds exactly to the on-site interactions introduced in Eq.~\eqref{onsite_interactions}. In Table~\ref{mapping}, we summarize the mapping of the phonon Hamiltonian onto the original periodically driven tunneling Hamiltonian of Sec.~\ref{pat_general}. As discussed there, since this interaction  is purely local, it shall not be modified by the scheme of PAT. Therefore, the effective phonon Hamiltonian becomes $H_{\rm eff}=K_{\rm eff}+V_{\rm eff}$, where 
\begin{equation}
\label{peierls_phonons}
\begin{split}
K_{\rm eff}&=\sum_{\alpha}\sum_{{\bf i}>{\bf j}}\tilde{J}_{{\rm d};{\bf i}{\bf j}}^{\alpha}\ee^{ie^*\int_{{\bf j}}^{{\bf i}}{\rm d}{\bf r}\cdot {\bf A}_{\rm s}}b_{\alpha,{\bf i}}^{\dagger}b_{\alpha,{\bf j}}^{\phantom{\dagger}}+\text{H.c.}\\
V_{\rm eff}&=\sum_{\ii,\alpha,\gamma}\tilde{U}_{\ii,\alpha\gamma}b_{\alpha,\ii}^{\dagger}b_{\gamma,\ii}^{\dagger}b_{\gamma,\ii}^{\phantom{\dagger}}b_{\alpha,\ii}^{\phantom{\dagger}},
\end{split}
\end{equation}
where the tunneling  is $\tilde{J}_{{\rm d};{\bf i}{\bf j}}^{\alpha}=J_{{\rm t};{\bf i}{\bf j}}^{\alpha}\mathcal{F}_{f({\bf i},{\bf j})}(\eta_{\rm d},\eta_{\rm d},\Delta\phi_{{\bf i}{\bf j}})$, and the gauge field can be expressed as  ${\bf A}_{\rm s}({\bf x})=-B_0y{\bf e}_1$, such that $B_0=(r\phi_2/e^*d_1d_2)$. Equation~\eqref{peierls_phonons}  incorporates the central result of this section, which shows that the PAT in microtrap arrays leads to a   quantum simulator of a long-range Bose-Hubbard model~\cite{bhm} under additional gauge fields~\cite{bhm_gauge}.

\begin{table}
   \caption{Mapping of microtrap phonons onto the PAT model}
 \begin{ruledtabular}
  \begin{tabular}{llllllll}
Eqs.~\eqref{tight_binding_general},\eqref{onsite_interactions}:& $a_{\sigma,\ii}^{\phantom{\dagger}}$ & $\sigma$& $J_{{\rm t};{\bf i}{\bf j}}^{\sigma}$ & $\omega_{\rm d}$ &$ \eta_{\rm d}\omega_{\rm d}$ & $\phi_{\bf i}$ & $\tilde{U}_{\sigma\sigma'}$\\
  \hline
Eqs.~\eqref{tight_binding_bis},\eqref{onsite_interactions_bis}: & $b_{\alpha,\ii}^{\phantom{\dagger}}$ &  $\alpha$ & $J_{{\rm c};{\bf i}{\bf j}}^{\alpha}$ & $\omega_{\rm L}$ & $-\Omega_{\rm L}\eta_{\alpha}^2$ & $\Delta\bf{k}\cdot {\bf r}_{\ii}^0$& $\half\tilde{\Omega}_{\rm L}\tilde{\eta}_{\alpha}^2\tilde{\eta}_{\gamma}^2\cos\tilde{\phi_{\ii}}$\\ 
  \end{tabular}
   \end{ruledtabular} 
   \label{mapping}
   \end{table}

There are several interesting regimes for this quantum simulator. In the non-interacting limit for a square microtrap array, it yields a bosonic {\it dipolar} version of the so-called {\it Azbel-Harper-Hofstadter model}~\cite{hofstadter_model}. The nearest-neighbor model has been studied thoroughly during the last decades, and contains several interesting properties that range from its relation to topological numbers, to the fractal and self-similar properties of its energy spectrum and wavefunctions,  the existence of gapless edge excitations, or the so-called $\pi$-flux phases~\cite{hofstadter_properties}. The addition of the long-range dipolar tunnelings introduces a new feature in the model that, to the best of our knowledge, has not been studied previously and may modify the above  phenomena. Besides, most of these effects rely on a magnetic flux per plaquette on the order of the  flux quantum, which cannot be achieved in solid-state materials assuming realistic magnetic fields. On the contrary,  our proposal  has the potential to reach these regimes since it is non-perturbative and the flux can attain arbitrary values $\phi_{\circlearrowleft}\in[0,2\pi]$.

Since it is possible to build any desired microtrap geometry~\cite{schmied_micro}, our quantum simulator can explore the physics of {\it bosonic ladders} subjected to synthetic gauge fluxes. Besides, the capability of tuning the fluxes, together with the independent control of the tunneling strength along/across the ladder rungs, dives into the phenomena of {\it flat-band physics} and {\it edge states}~\cite{flat_band_ladders}, which is typical for fermionic topological insulators that break the time-reversal symmetry~\cite{ti_review}.

Another interesting regime corresponds to $|J_{\rm d;{\bf i}{\bf j}}^{\alpha}|\approx \tilde{U}_{{\bf i},\alpha\gamma}$, where the interactions compete with the kinetic energy and induce  strong correlations in the Bose-Hubbard model~\cite{bhm}. With respect to the neutral-atom realizations~\cite{bhm_atoms}, the phonon model includes the effects of longer-range tunnelings, and the possibility to address site-dependent interactions. Besides,  thanks to the tunability of  the functions $\mathcal{F}_{f({\bf i},{\bf j})}$ [Fig.~\ref{fig_amplitude}], the strongly-correlated regime can  be reached, in principle,  regardless of how small the on-site interactions are. Let us note, however, that there is a fundamental limit to this approach, which is imposed by external sources of decoherence and heating. Accordingly, the dynamics must always be faster than the time-scale imposed by these sources of noise. 

For vibrational modes transverse to the microtrap array $\alpha=z$,  the condition $\tilde{\Omega}_{\rm L}\tilde{\eta}_{\alpha}\ll \omega_{\alpha}$ can be relaxed for ions  in the node of the standing wave. It then suffices to set $\tilde{\Omega}_{\rm L}\tilde{\eta}^2_{\alpha}\ll \omega_{\alpha}$, which allows us to reach interaction strengths in the $\tilde{U}\sim$1-10 kHz-regime, which are directly on the order of the bare tunnelings. Finally, including the gauge fields in this strongly-correlated regime further enhances the versatility of our quantum simulator. Even if the particles are  bosonic and the interactions are local, one can target {\it fractional quantum Hall states}, and {\it composite-fermion fluids}~\cite{bhm_gauge}.

\subsection{Non-Abelian synthetic gauge fields}
\label{non_abelian}

The synthetic gauge fields  discussed so far~\eqref{peierls_phonons} correspond to  standard electromagnetism. In this theory, they are formally introduced to restore the invariance   with respect to local unitary transformations  in the gauge group U(1).  A natural question that arises is whether these synthetic fields can be generalized to different gauge groups, possibly non-Abelian ones~\cite{non_abelian}. In this case, the local unitary also acts on some additional degree of freedom, which we shall refer to as  the flavor.  Here, we show that it is possible to realize such scenarios by exploiting two or three orthogonal directions of vibration  as the different flavors of the non-Abelian theory.

We first exploit two main axes of vibration $\alpha=x,y$ within the plane defined by the microtrap array,  the corresponding trapping frequencies being different $\omega_x\neq\omega_y$. In the regime where the bare tunneling strength is smaller than the frequency difference, these two directions are uncoupled (see Eq.~\eqref{tight_binding_bis}). The main idea  is to use two gradients of opposite sign for each degree of freedom, and two separate periodic drivings of the same frequency but of different amplitudes 
\begin{equation}
\label{su2_gradients}
\begin{split}
\omega_{x,{\bf i}}&=\omega_x+\Delta\omega i_1+\eta_{{\rm d}x}\omega_{\rm d}\cos(\omega_{\rm d}t+\phi_{\bf i}),\\
\omega_{y,{\bf i}}&=\omega_y-\Delta\omega i_1+\eta_{{\rm d}y}\omega_{\rm d}\cos(\omega_{\rm d}t+\phi_{\bf i}).\\
\end{split}
\end{equation}
According to Eq.~\eqref{periodic_driving}, this modulation of the trap frequencies  can be obtained from a single Raman-beam scheme, such that the corresponding wavevector has a component along both  axes $\eta_{{\rm d}x}=-\Omega_{\rm L}\eta_x^2$, $\eta_{{\rm d}y}=-\Omega_{\rm L}\eta_y^2$.  By repeating the analysis of Sec.~\ref{pat_general}, and considering the same type of conditions, we reveal that the dressed tunneling strengths~\eqref{dressed_coupling} must be modified for each of the vibrational axes
     \begin{equation}
   \label{dressed_coupling_ions}
J_{{\rm d};{\bf i}{\bf j}}^{\alpha}= \tilde{J}_{{\rm d};{\bf i}{\bf j}}^{\alpha}\ee^{-i\frac{f_{\alpha}({\bf i},{\bf j})}{2}(\phi_{{\bf i}}+\phi_{\bf j})},\hspace{1ex}\tilde{J}_{{\rm d};{\bf i}{\bf j}}^{\alpha}=J_{{\rm t};{\bf i}{\bf j}}^{\alpha}\mathcal{F}_{f_{\alpha}({\bf i},{\bf j})}(\eta_{\rm d\alpha},\eta_{\rm d\alpha},\Delta\phi_{{\bf i}{\bf j}})
   \end{equation}
   where $f_{\alpha}({\bf i},{\bf j})=r_{\alpha}(i_1-j_1)$, and one must introduce the following axis-dependent parameter $r_x=r,r_y=-r$. It is then straightforward to rewrite the effective Hamiltonian according to a generalized Peierls substitution
\begin{equation}
\begin{split}
K_{\rm eff}=\sum_{{\bf i}>{\bf j}}\tilde{J}_{{\rm d};{\bf i}{\bf j}}^{x}\ee^{ie^*\int_{{\bf j}}^{{\bf i}}{\rm d}{\bf r}\cdot {\bf A}^x_{\rm s}}b_{x,{\bf i}}^{\dagger}b_{x,{\bf j}}^{\phantom{\dagger}}+\tilde{J}_{{\rm d};{\bf i}{\bf j}}^{y}\ee^{ie^*\int_{{\bf j}}^{{\bf i}}{\rm d}{\bf r}\cdot {\bf A}^y_{\rm s}}b_{y,{\bf i}}^{\dagger}b_{y,{\bf j}}^{\phantom{\dagger}}+\text{H.c.},
\end{split}
\end{equation}
where the synthetic gauge potential now depends on the related vibrational axis, and is equivalent to ${\bf A}^x_{\rm s}({\bf r})=-{\bf A}^y_{\rm s}({\bf r})=-B_0y {\bf e}_1$, such that $B_0=r\phi_2/e^*d_1d_2$. Let us introduce now a bosonic spinor field operator $\Psi_{\bf i}=(b_{x,\ii},b_{y,\ii})^{t}$ to describe the phonon fields corresponding to vibrations in each direction. The kinetic part can be written
\begin{equation}
\begin{split}
K_{\rm eff}
=
\sum_{{\bf i}>{\bf j}}
\Psi_{{\bf i}}^{\dagger}
K_{{\rm d};{\bf i}{\bf j}}
\ee^{ie^*\int_{{\bf j}}^{{\bf i}}{\rm d}{\bf r}\cdot 
{\bf A}^{{\rm na}}_{\rm s}}\Psi_{{\bf j}}^{\phantom{\dagger}}+\text{H.c.},
\label{Keff}
\end{split}
\end{equation}
where $K_{{\rm d};{\bf i}{\bf j}}$ is a matrix that describes the vibrational couplings for each spinor component,
\begin{equation}
K_{{\rm d};{\bf i}{\bf j}} = 
\left(
\begin{array}{cc}
\tilde{J}^x_{{\rm d};{{\bf i},{\bf j}}} & 0 \\
0 & \tilde{J}^y_{{\rm d};{{\bf i},{\bf j}}}
\end{array}
\right) ,
\end{equation}
and ${\bf A}^{\rm na}_{\rm s}= -B_0y\tau_{z}{\bf e}_1$ is a {\it SU(2) non-Abelian gauge field}, that we write in terms of  a Pauli matrix, $\tau_z$, acting on the vibrational index (i.e. flavor index).  The associated magnetic field corresponds to ${\bf B}^{\rm na}_{\rm s}=B_0\tau_{z}{\bf e}_z$, so that each flavor is subjected to an opposite flux piercing the lattice. For fermions, these types of gauge fields give rise to the so-called quantum spin Hall effect~\cite{qsh}, which is the prototype of time-reversal preserving topological insulators in two-dimensions~\cite{ti_review}.
Let us note that Eq. (\ref{Keff}), together with the arguments presented in Eq. (\ref{flux.2}) imply that each flavor around a plaquette accumulates a non-Abelian Aharanov-Bohm phase that is governed solely by the SU(2) gauge field ${\bf A}^{\rm na}_{\rm s}$. However, in addition to that, each flavor $x$, $y$, is subjected to different tunnelings, $\tilde{J}^x_{{\rm d};{{\bf i},{\bf j}}}$, $\tilde{J}^y_{{\rm d};{{\bf i},{\bf j}}}$, such that an spin-orbit coupling is superimposed to the non-Abelian gauge. The latter effect may enrich the dynamics with respect to the usual situation in SU(2) gauge  theories.

We remark that the above synthetic gauge field~\eqref{Keff} is only a particular type of non-Abelian SU(2) gauge fields. In order to consider more general fields, it is possible to exploit the quadratic terms of the Coulomb interaction~\eqref{coulomb_mixing} that mix the vibrational modes along $x,y$. By setting the frequency of the periodic driving to account for both the frequency difference ($\omega_x\neq\omega_y$) and the particular gradient, the dressed tunnelings would also involve a change of the flavor index, yileding thus more general gauge fields in the group SU(2).

\subsection{Spin-mediated disordered  Hamiltonians}
\label{disorder}

The properties of solids usually differ from those of perfectly periodic crystals. In realistic samples, there is a certain amount of disorder in the form of impurities, dislocations, or vacancies, which may alter dramatically the  properties of the solid. The study of disorder in solids is an active and mature field of condensed matter~\cite{ziman_disorder}, where the system Hamiltonians are usually modeled as stochastic operators. Here,  the randomness is due to a statistical description of the disordered degrees of freedom. To incorporate such a randomness in a quantum simulator, which by definition should be an extremely clean and controllable  setup, one can exploit the quantum parallelism  by an auxiliary degree of freedom~\cite{qs_random}.

In principle, some randomness could be introduced by randomly varying the lattice constant and the trapping frequencies within the microtrap array. In the limit of  large arrays, these would lead to an off-diagonal bond disorder and diagonal site disorder, respectively. In this subsection, we elaborate on two alternative  directions  to widen the applicability of the phonon-based quantum simulator by introducing disorder regardless of the size of the microtrap array.  In both cases, we shall make use of the electronic energy levels of the ion, $\{\ket{\uparrow_{\bf i}},\ket{\downarrow_{\bf i}}\}$,  to introduce randomness in the phonon Hamiltonian and  mimic the effects of disorder. We emphasize that this  setup shall allow us to control the two usual types of disorder independently, namely, diagonal and off-diagonal disorder. 

\subsubsection{Off-diagonal bond disorder}

\begin{figure}
\centering
\includegraphics[width=1\columnwidth]{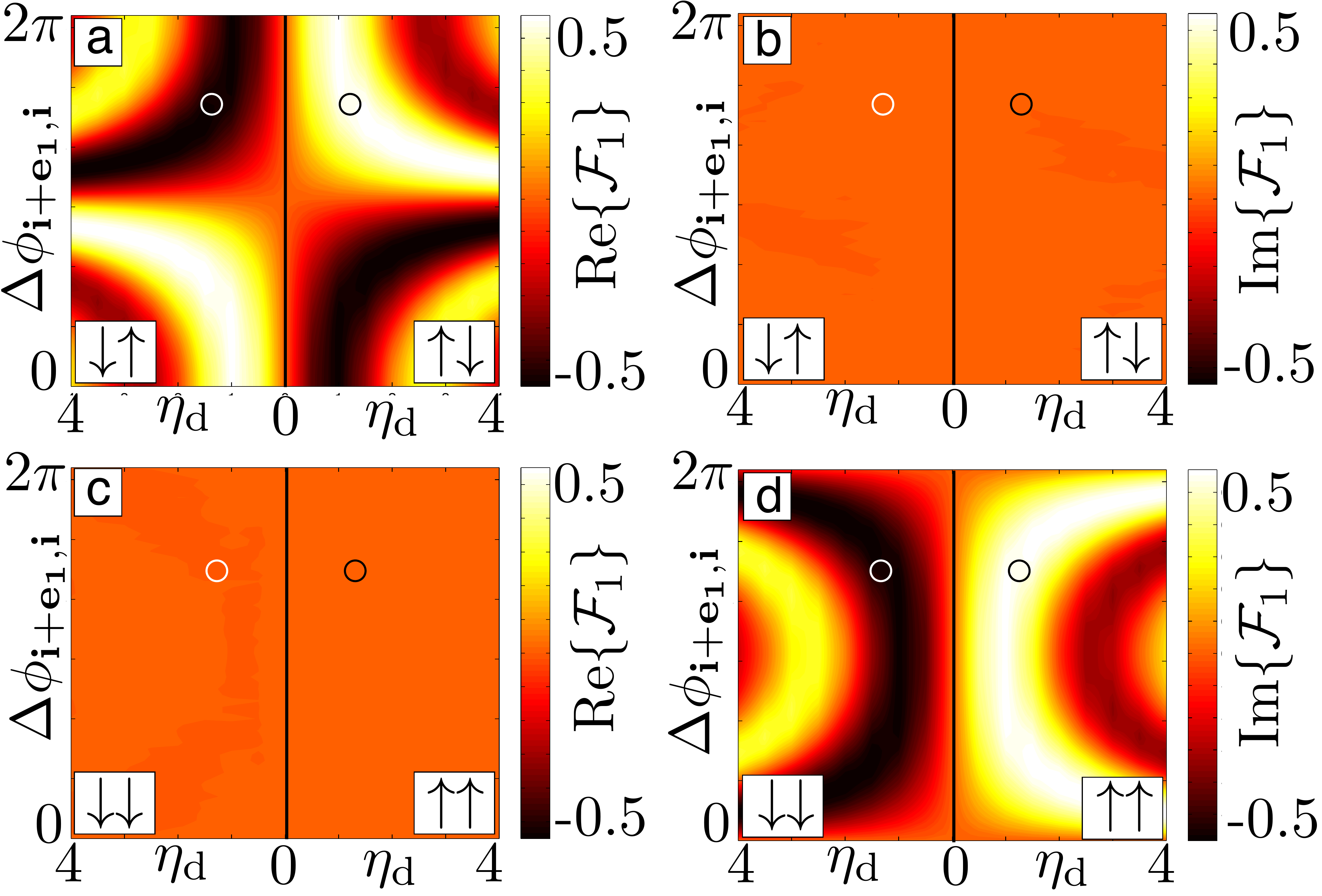}
\caption{ {\bf Spin-dependent modulation of the tunneling:}  Contour plot of the real and imaginary parts of the modulation amplitude $\mathcal{F}_{r(i_1-j_1)}$ for the tunneling between nearest-neighbors  ${\bf i}\to{\bf j}$, such that $i_1=j_1+1$, as a function of the driving parameters $\eta_{\rm d},\Delta\phi_{{\bf i}{\bf j}}$ for $r=1$. {\bf (a,b)} Tunneling in the configuration of anti-parallel spins $\sigma_{\bf i},\sigma_{{\bf j}}\in\{\uparrow\downarrow,\downarrow\uparrow\}$, {\bf (c,d)} Tunneling in the configuration of parallel spins $\sigma_{\bf i},\sigma_{{\bf j}}\in\{\uparrow\uparrow,\downarrow\downarrow\}$. Also shown in black-white circles the region of interest for $\eta_{\rm d}\approx 1,\Delta\phi_{\rm i j}=3\pi/2$.}
\label{fig_amplitude_spins}
\end{figure}

 We discuss how to induce randomness on the phonon tunneling  by a slight modification of Eq.~\eqref{periodic_driving}, so that the periodic driving will be responsible for both the disorder and the synthetic gauge fields. In the derivation of the driving, we assumed that a proper choice of the laser intensities, detunings, and polarizations would lead us  from Eq.~\eqref{raman} to the desired expression~\eqref{stark_shift}.
This condition  can be modified so that the periodic driving becomes spin dependent. In fact, by setting  $\Omega_{\rm L}= \big(\Omega^{(1)}_{\rm ag}\big)^*\Omega^{(2)}_{\rm ag}/2\Delta=-\big(\Omega^{(1)}_{\rm ae}\big)^*\Omega^{(2)}_{\rm ae}/2\Delta$, one obtains
\begin{equation}
\label{spin_dependent_stark_shift}
H_{\rm L}=\half\Omega_{\rm L}\sum_{\bf i}\sigma_{\bf i}^z\ee^{i(\Delta {\bf k}\cdot{\bf r}_{\bf i}-\omega_{\rm L}t)}+\text{H.c.},
\end{equation}
where we have introduced $\sigma_{\bf i}^z=\ket{\uparrow_{\bf i}}\bra{\uparrow_{\ii}}-\ket{\downarrow_{\ii}}\bra{\downarrow_{\ii}}$. This expression corresponds to a differential Stark shift between the two electronic levels. In the regime where $\omega_{\rm L}\ll \omega_{\alpha}$, this term generalizes Eq.~\eqref{periodic_driving} to the following spin-dependent  driving of the trapping frequencies 
\begin{equation}
\label{spin_periodic_driving}
H_{\rm L}\approx-\Omega_{\rm L}\sum_{\alpha,{\bf i}} \eta_{\alpha}^2\cos(\Delta{\bf k}\cdot{\bf r}_{\bf i}^0-\omega_{\rm L}t)\sigma_{\ii}^z\hspace{0.2ex}b^{\dagger}_{\alpha,{\bf i}}b_{\alpha,{\bf i}}^{\phantom{\dagger}},
\end{equation}
where the same constraints~\eqref{constraints} over the system parameters must be fulfilled. Since no other spin operators are involved in the Hamiltonian, one can treat $\sigma_{\bf i}^z$ as $c$-numbers $\sigma_{\ii}\in\{-1,1\}$, and carry out the same analysis made in Sec.~\ref{pat_general}. In fact, one should simply modify Hamiltonian~\eqref{peierls_phonons} to $H_{\rm eff}=\sum_{\{\boldsymbol{\sigma}\}}H_{\rm eff}(\{\boldsymbol{\sigma}\})\ket{\{\boldsymbol{\sigma}\}}\bra{\{\boldsymbol{\sigma}\}}$, where $H_{\rm eff}(\{\boldsymbol{\sigma}\})=K_{\rm eff}(\{\boldsymbol{\sigma}\})+V_{\rm eff}$, such that 
\begin{equation}
\label{bond_disorder}
K_{\rm eff}(\{\boldsymbol{\sigma}\})=\sum_{\alpha}\sum_{{\bf i}>{\bf j}}\tilde{J}_{{\rm d};{\bf i}{\bf j}}^{\alpha}(\sigma_{\bf i},\sigma_{\bf j})\ee^{ie^*\int_{{\bf j}}^{{\bf i}}{\rm d}{\bf r}\cdot {\bf A}_{\rm s}}b_{\alpha,{\bf i}}^{\dagger}b_{\alpha,{\bf j}}^{\phantom{\dagger}}+\text{H.c.},
\end{equation}
and  the tunneling amplitudes  account for the different spin configurations $\tilde{J}_{{\rm d};{\bf i}{\bf j}}^{\alpha}(\sigma_{\bf i},\sigma_{\bf j})={J}_{{\rm t};{\bf i}{\bf j}}^{\alpha}\mathcal{F}_{f({\bf i},{\bf j})}(\eta_{\rm d}\sigma_{\ii},\eta_{\rm d}\sigma_{{\bf j}},\Delta\phi_{{\bf i}{\bf j}})$.
In Fig.~\ref{fig_amplitude_spins}, we represent the dressed tunneling amplitudes between nearest-neighbors along the direction of the gradient. It can be observed that depending on the spin state, one obtains different values. In particular, for $\eta_{\rm d}\approx 1$ and $\Delta\phi\approx 3\pi/2$, the tunneling amplitude can attain four possible values
\begin{equation}
\label{hopping_values}
 \mathcal{F}_{1}^{\uparrow\downarrow}=- \mathcal{F}_{1}^{\downarrow\uparrow}=i \mathcal{F}_{1}^{\downarrow\downarrow}=-i \mathcal{F}_{1}^{\uparrow\uparrow}\approx 0.5.
\end{equation}
   If we now consider the following initial state $\rho_0=\ket{\Psi_{\rm s}}\bra{\Psi_{\rm s}}\otimes\rho_{\rm ph}^0$, where $\ket{\Psi_{\rm s}}=\sum_{\{\boldsymbol{\sigma}\}}c_{\{\boldsymbol{\sigma}\}}\ket{\{\boldsymbol{\sigma}\}}$  and $\rho_{\rm ph}^0$ are arbitrary spin and phonon states, its time-evolution is 
\begin{equation}
\label{disorder_evolution}
\rho_{\rm ph}(t)={\rm tr}_s\{\rho(t)\}=\sum_{\{\boldsymbol{\sigma}\}}|c_{\{\boldsymbol{\sigma}\}}|^2\ee^{-i  H_{\rm eff}(\{\boldsymbol{\sigma}\})t}\rho_{\rm ph}^0\ee^{+i  H_{\rm eff}(\{\boldsymbol{\sigma}\})t}.
\end{equation}
Note that due to the superposition principle of quantum mechanics, the phonon state explores simultaneously the different tunneling paths ${\bf i}\to  J({\sigma_{\bf i}\sigma_{\bf j}})\to{\bf j}$ with a probability that depends on the initial spin configuration $p_{\{\boldsymbol{\sigma}\}}=|c_{\{\boldsymbol{\sigma}\}}|^2$. In fact, the measurement of a phonon observable $O_{\rm ph}$ corresponds directly to the statistical average over all such tunneling paths
\begin{equation}
\label{statistical_avg}
\begin{split}
\langle O_{\rm ph}(t)\rangle&=\sum_{\{\boldsymbol{\sigma}\}}p_{\{\boldsymbol{\sigma}\}}{\rm tr}\{O_{\rm ph}\ee^{-i  H_{\rm eff}(\{\boldsymbol{\sigma}\})t}\rho_{\rm ph}^0\ee^{+i  H_{\rm eff}(\{\boldsymbol{\sigma}\})t}\}\\
&=\sum_{\{\boldsymbol{\sigma}\}}p_{\{\boldsymbol{\sigma}\}}\langle O^{\{\boldsymbol{\sigma}\}}_{\rm ph}(t)\rangle,
\end{split}
\end{equation}
which can be understood as the average over all possible realizations of the bond disorder. Therefore, our phonon-based simulator can explore the physics of interacting disordered bosons in a lattice pierced by an external magnetic field. 

\subsubsection{ Diagonal site disorder}

 We now address a scheme to introduce randomness in the on-site energies~\cite{anderson}. The origin of such terms is independent of the periodic driving, and requires a pair of additional laser beams in the Raman configuration~\eqref{laser_ion}. In order to distinguish them from the previous laser beams, we will be denote them  with an overbar. The main difference with respect to the previous case is that  the beatnote is tuned close to the transition between the internal states $\bar{\omega}_{\rm L}=\bar{\omega}_1-\bar{\omega}_2\approx\omega_0$. In this case, the laser-ion Hamiltonian becomes
\begin{equation}
\label{sideband}
\bar{H}_{\rm L}=\half\bar{\Omega}_{\rm L}\sum_{\bf i}\sigma_{\bf i}^+\ee^{i(\Delta {\bf k}\cdot{\bf r}_{\bf i}-(\omega_0-\bar{\omega}_{\rm L})t)}+\text{H.c.},
\end{equation}
where we have introduced $\sigma_{\ii}^+=\ket{\uparrow_{\ii}}\bra{\downarrow_{\bf i}}$. In analogy to the derivation of the periodic driving,  we express the position in terms of the local phonon operators, and  expand the Hamiltonian for a small Lamb-Dicke parameter. If the laser beatnote is tuned as follows $\bar{\omega}_{\rm L}\approx \omega_{0}-\omega_{\alpha}$, one obtains the so-called red-sideband excitation
\begin{equation}
\begin{split}
\bar{H}_{\rm L}&\approx i \half\bar{\Omega}_{\rm L}\sum_{\alpha,\bf i}\eta_{\alpha}\ee^{i\Delta{\bf k}\cdot {\bf r}_{\bf i}^0}\sigma_{\bf i}^+b_{\alpha,{\bf i}}^{\phantom{\dagger}}\ee^{-i\bar{\delta}_{\rm L}t}+\text{H.c.},
\end{split}
\end{equation}
where the bare detuning is $\bar{\delta}_{\rm L}=\omega_0-\omega_{\alpha}-\bar{\omega}_{\rm L}\ll\omega_{\alpha}$, and we assume that $\bar{\Omega}_{\rm L}\ll\omega_{\alpha}$ in order to neglect the remaining terms of the Taylor expansion. In the regime where the laser beams are weak enough $\bar{\Omega}_{\rm L}\eta_{\alpha}\ll \bar{\delta}_{\rm L}$, it is possible to find the following laser-ion Hamiltonian in perturbation theory
\begin{equation}
\label{phonon_stark}
\bar{H}_{\rm L}\approx\sum_{\alpha,\ii}\epsilon_{\alpha}\sigma_{\ii}^zb_{\alpha,\ii}^{\dagger}b_{\alpha,\ii}^{\phantom{\dagger}},\hspace{1ex}\textstyle{\epsilon_{\alpha}=\frac{\bar{\Omega}_{\rm L}^2\eta_{\alpha}^2}{4\bar{\delta}_{\rm L}}}.
\end{equation} 
This term arises due to a second-order process where a phonon is virtually excited and reabsorbed by a single ion, and leads to a differential Stark shift of the atomic levels that depends on the number of phonons. When incorporated to the effective description of the PAT Hamiltonian~\eqref{peierls_phonons}, it modifies the Kinetic energy term. If one is interested in the time span $t\approx 1/J_{\rm c; {\bf i}{\bf j}}$, the dynamics of the spins can be safely ignored, and $\sigma_{\bf i}^z$ can be treated once more as $c$-numbers $\sigma_{\ii}\in\{-1,1\}$. We note that the typical time-scales for the spin flip-flop dynamics would be $J_{{\rm s};{\bf i}{\bf j}}\approx$10-100 Hz, whereas thevibrational couplings lie in the $J_{\rm c; {\bf i}{\bf j}}\approx$1-10 kHz. Accordingly, the Kinetic energy term of Eq.~\eqref{peierls_phonons} must be modified to $H_{\rm eff}(\{\boldsymbol{\sigma}\})=K_{\rm eff}(\{\boldsymbol{\sigma}\})+V_{\rm eff}$, where
\begin{equation}
\label{site_disorder}
K_{\rm eff}(\{\boldsymbol{\sigma}\})=\sum_{\alpha}\sum_{{\bf i}>{\bf j}}\tilde{J}_{{\rm d};{\bf i}{\bf j}}^{\alpha}\ee^{ie^*\int_{{\bf j}}^{{\bf i}}{\rm d}{\bf r}\cdot {\bf A}_{\rm s}}b_{\alpha,{\bf i}}^{\dagger}b_{\alpha,{\bf j}}^{\phantom{\dagger}}+\textstyle{\frac{\epsilon_{\alpha}}{2}}\sigma_{\bf i}b_{\alpha,{\bf i}}^{\dagger}b_{\alpha,{\bf i}}^{\phantom{\dagger}}+\text{H.c.}
\end{equation}
In this situation, it is not the tunneling strengths which depend upon the spin state~\eqref{bond_disorder}, but rather the on-site energies. Notice that the strength of these on-site energies must be much smaller that the trapping gradient $\epsilon_{\alpha}\ll\Delta\omega_{\alpha}$ in order not to affect  the PAT scheme.
According to Eq.~\eqref{statistical_avg},  the system explores simultaneously all possible values of the on-site energies with probabilities that depend on the spin configurations of the initial state. Therefore, the measurement of phonon observables yields directly the statistical average over all possible realization of  $\epsilon_{j}(\{\boldsymbol{\sigma}\})\in\{-\epsilon_{\alpha},\epsilon_{\alpha}\}$ with a probability distribution  $p_{\{\boldsymbol{\sigma}\}}=|c_{\{\boldsymbol{\sigma}\}}|^2$.

Let us now comment on the extended possibilities of our quantum simulator due to the engineered disorder in Eqs.~\eqref{bond_disorder} and~\eqref{site_disorder}. The diagonal site disorder leads to the well-known {\it Anderson localization} in the non-interacting limit~\cite{and_localization}. This phenomenon is due to the interference of the different paths associated to the scattering of the particles from the random on-site fluctuations, and gives rise to exponentially-localized wavefunctions and absence of diffusion. The combination of Anderson Localization with strong interactions, which is also a well-studied problem~\cite{bhm}, leads to interesting insulating, yet gapless, phases such as the {\it Bose glass}. 
In the case of strong bond disorder, a different gapless insulator known as the {\it Mott glass} arises, which consists of disconnected superfluid regions of random size~\cite{bhm_bond_disorder}. Besides, our quantum simulator has the potential of combining both bond and site disorder, and tuning them independently, which may pave the way towards other exotic insulating phases. In addition to the aforementioned Bose and Mott glasses, the simulator can explore the {\it random-singlet glass} where the bosons form delocalized random pairs~\cite{bhm_both_disorder}. The possibility to explore higher-dimensions, longer-range tunnelings, and the effect of synthetic gauge fields makes our scheme a very versatile tool.

\subsection{ Decorated synthetic gauge flux lattices }
\label{flux_lattices}

In this section, we describe an additional feature of our quantum simulator: the possibility to decorate the lattice with any desired  pattern of synthetic fluxes. It is thus possible to engineer highly inhomogeneous synthetic gauge fields, even reaching inhomogeneities at the unit-cell limit. The idea is to use the spins to decorate the array with different fluxes by exploiting the differential phonon-dependent Stark shift~\eqref{phonon_stark}. 

\begin{figure*}
\centering
\includegraphics[width=1.9\columnwidth]{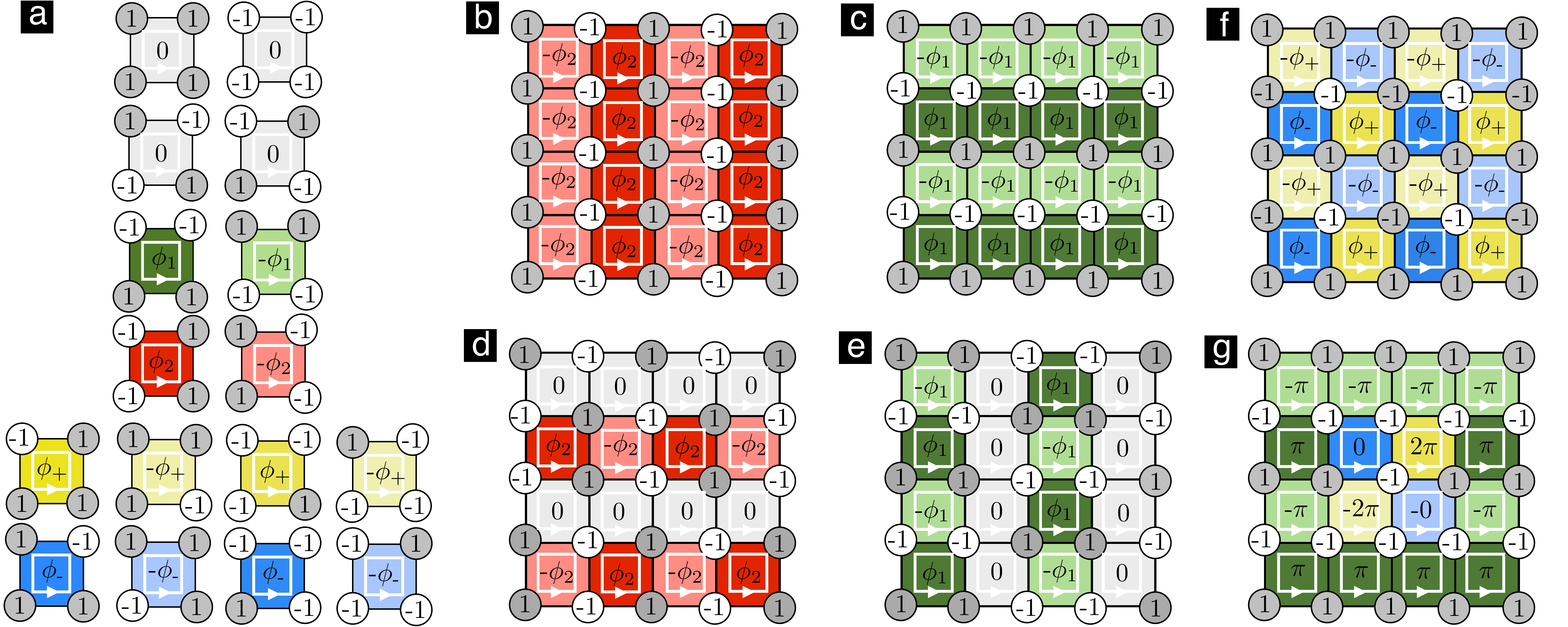}
\caption{ {\bf Decorated synthetic flux lattices:} {\bf (a)} Unit cell for a square array of microtraps with all the possible configurations of spin states $\sigma_{\bf i}\in\{1,-1\}$, which are represented by circles at the vertices of the cell. According to the spin-dependent enclosed flux, we have a total of 9 different plaquettes that can act as tiles to construct the decorated flux lattices. {\bf (b-c)} Staggered flux lattices. {\bf (d-e)} Staggered flux lattices with alternating strings of zero-flux plaquettes. {\bf (f)} Tetra-flux checkerboard lattice. {\bf (g)} $\pi$-flux lattice with a localized defect consisting of a zero-flux plaquette. }
\label{fig_fluxes}
\end{figure*}

We consider a situation where the gradient of the trapping frequencies vanishes $\Delta\omega_{\alpha}=0$, so that the regime is different  from that of the site-disorder case considered above $\epsilon_{\alpha}\ll\Delta\omega_{\alpha}$.  When the periodic driving frequency~\eqref{periodic_driving} and the Stark shift~\eqref{phonon_stark} fulfill the resonance condition $\omega_{\rm L}=2\epsilon_{\alpha}/r$, where $r$ is the integer representing the umber of photons involved in the PAT, the assisted tunneling will give rise to a different phase depending on the spin states of the two neighboring ions. In order to find the correct expression for the tunneling, we readdress the derivation of Sec.~\ref{pat_general} for this particular spin-dependent situation. Since we are interested in the phonon dynamics, the spins are effectively frozen, and we can consider  $\sigma_{\bf i}^z$ as $c$-numbers $\sigma_{\ii}\in\{-1,1\}$. Hence, the dressed tunneling~\eqref{dressed_coupling} must be modified as follows
 \begin{equation}
   \label{dressed_coupling_spin_bis}
J_{{\rm d};{\bf i}{\bf j}}^{\alpha}= \tilde{J}_{{\rm d};{\bf i}{\bf j}}^{\alpha}\ee^{-i\frac{f(\sigma_{\bf i},\sigma_{\bf j})}{2}(\phi_{{\bf i}}+\phi_{\bf j})},\hspace{1ex}\tilde{J}_{{\rm d};{\bf i}{\bf j}}^{\alpha}=J_{{\rm t};{\bf i}{\bf j}}^{\alpha}\mathcal{F}_{f(\sigma_{\bf i},\sigma_{\bf j})}(\eta_{\rm d},\eta_{\rm d},\Delta\phi_{{\bf i}{\bf j}})
   \end{equation}
   where we have introduced $f(\sigma_{\bf i},\sigma_{\bf j})=r(\sigma_{\bf i}-\sigma_{\bf j})/2$. By considering the possible spin configurations, we find
    \begin{equation}
   \label{dressed_coupling_spin}
\frac{J_{{\rm d};{\bf i}{\bf j}}^{\alpha}(\{\boldsymbol{\sigma}\})}{J_{{\rm t};{\bf i}{\bf j}}^{\alpha}}=  \hspace{-0.5ex}\left\{\begin{array}{l}\mathcal{F}_{r}(\eta_{\rm d},\eta_{\rm d},\Delta\phi_{{\bf i}{\bf j}})\ee^{-i\frac{r}{2}(\phi_{{\bf i}}+\phi_{\bf j})}
,\hspace{0.5ex} \phantom{-} \text{if} \hspace{0.25ex} \sigma_{\bf i}=-\sigma_{\bf j}=1
 \\ \mathcal{F}_{-r}(\eta_{\rm d},\eta_{\rm d},\Delta\phi_{{\bf i}{\bf j}})\ee^{+i\frac{r}{2}(\phi_{{\bf i}}+\phi_{\bf j})}
 ,\hspace{0.5ex} \text{if} \hspace{0.5ex} \sigma_{\bf j}=-\sigma_{\bf i}=1
 \\ \mathcal{F}_{0}(\eta_{\rm d},\eta_{\rm d},\Delta\phi_{{\bf i}{\bf j}}),\phantom{\ee^{+i\frac{r}{2}(\phi_{{\bf i}}+\phi_{\bf j})}}
 \hspace{1.5ex} \text{if} \hspace{0.5ex} \sigma_{\bf j}=\sigma_{\bf i}=\pm 1.
\end{array}\right.
   \end{equation}
   Interestingly, the phase of the tunneling between  sites ${\bf j}\to{\bf i}$ depends on their internal spin state  $\sigma_{\bf j}\to\sigma_{\bf i}$. If their spins are parallel $\sigma_{\bf i}=\sigma_{\bf j}=\pm 1$, the phase vanishes and thus the tunneling does not contribute to the synthetic gauge field. Conversely, when the spins are anti-parallel,  the phases contribute to the gauge fluxes with a sign that depends on the particular spin ordering. We have calculated  the consecutive  phonon tunneling   around a square plaquette ${\bf r}_{\bf i}^0\to{\bf r}_{\bf j}^0\to {\bf r}_{\bf k}^0\to{\bf r}_{\bf l}^0\to{\bf r}_{\bf i}^0$ for all the possible spin configurations 
\begin{equation}
W^{(1)}_{\circlearrowleft}=J^{\alpha}_{{\rm d};{\bf i}{\bf l}}(\sigma_{\bf i},\sigma_{\bf l})J^{\alpha}_{{\rm d};{\bf l}{\bf k}} (\sigma_{\bf l},\sigma_{\bf k}) J^{\alpha}_{{\rm d};{\bf k}{\bf j}}(\sigma_{\bf k},\sigma_{\bf j})J^{\alpha}_{{\rm d};{\bf j}{\bf i}} (\sigma_{\bf j},\sigma_{\bf i}).
\end{equation}
All the possible encircled fluxes $W^{(1)}_{\circlearrowleft}=|W^{(1)}_{\circlearrowleft}|\ee^{i\phi_{\circlearrowleft}(\{\boldsymbol{\sigma}\})}$ for $r=1$ have been represented in Fig.~\ref{fig_fluxes}{\bf (a)}, where we observe that there are 9 possible fluxes out of the $2^4=16$ possible spin configurations. These fluxes correspond to $\phi_{\circlearrowleft}\in\{0,\pm \phi_1,\pm\phi_2,\pm\phi_+,\pm\phi_-\}$, where $\phi_{i}=\Delta k_id_i$, and $\phi_{\pm}=(\phi_1\pm\phi_2)/2$. Accordingly, we have 9 different tiles that can be used to decorate the underlying microtrap array with a particular distribution of fluxes. Let us emphasize that the particular distribution of tiles is completely determined by the spin state $\ket{\Psi_{\rm s}}=\ket{\sigma_1,\cdots,\sigma_{N}}$, which can be initialized at will in trapped-ion experiments. Let us also note that we could also exploit this result to introduce randomness in the gauge fields by considering a linear superposition of the spin states.

In Fig.~\ref{fig_fluxes}{\bf (b-c)}, we represent two-color flux lattices that correspond to a staggered magnetic field along both principal axes. In Fig.~\ref{fig_fluxes}{\bf (d-e)}, we represent three-color flux lattices where a staggered flux alternates with a vanishing flux. In Fig.~\ref{fig_fluxes}{\bf (f)}, we represent a checkerboard flux lattice, and finally in Fig.~\ref{fig_fluxes}{\bf (g)}, we represent a limiting case of a six-color flux lattice, which is very interesting from a physical point of view. By setting $\phi_1=\pi$, and $\phi_2=\pi$, the two fluxes $\pm \phi_1=\pm \pi=\pi{\rm mod}2\pi$ are equivalent, and lead to a homogeneous $\pi$-flux model in a square lattice. Additionally, the  four remaining fluxes vanish for this choice $\pm \phi_+=\pm\phi_-=0{\rm mod}2\pi$. Therefore they contribute with a local defect over the $\pi$-flux lattice, which could  bind excitations with {\it anyonic statistics} when the longer range tunnelings are taken into account~\cite{anyons_flux}. 

For each of the decorated flux lattices, it is possible to find a particular inhomogeneous synthetic gauge field  ${\bf A}_{\rm s}(\sigma_{\bf i},{\sigma_{\bf j}})$ so that the effective phonon Hamiltonian is rewritten in a standard  Peierls form 
\begin{equation}
K_{\rm eff}(\{\boldsymbol{\sigma}\})=\sum_{\alpha}\sum_{{\bf i}>{\bf j}}\tilde{J}_{{\rm d};{\bf i}{\bf j}}^{\alpha}(\sigma_{\bf i},\sigma_{\bf j})\ee^{ie\int_{{\bf j}}^{{\bf i}}{\rm d}{\bf x}\cdot {\bf A}_{\rm s}(\sigma_{\bf i},{\sigma_{\bf j}})}b_{\alpha,{\bf i}}^{\dagger}b_{\alpha,{\bf j}}^{\phantom{\dagger}}+\text{H.c.}.
\end{equation}
 The particular pattern of the spins will determine the inhomogeneous gauge field, and the way the lattice is decorated with fluxes. Let us also note that the complexity of Fig.~\eqref{fig_fluxes} will increase when the larger plaquettes due to long-range tunnelings are also taken into account.

\section{Summary and Outlook}
\label{summary}

We have introduced the two key ingredients to realize PAT experiments in micro-fabricated ion traps. The first ingredient is a gradient of the trapping frequencies achieved by the local control of the trap electrodes. The second corresponds to a Raman-beam configuration, which presents different regimes that provide (i) the periodic driving of the trapping frequencies, (ii) the on-site phonon-phonon interactions, (iii) the bond/site disorder, and also (iv) an exotic flux decoration of the microtrap array. We believe that such ingredients are at reach of current microtrap technology, and their correct combination will give raise to a very versatile quantum simulator for many-body bosonic models.  Such {\it photon-assisted-tunneling toolbox} for quantum simulations can be summarized in the following general Hamiltonian
$H_{\rm eff}=\sum_{\{\boldsymbol{\sigma}\}}(K_{\rm eff}(\{\boldsymbol{\sigma}\})+V_{\rm eff})\ket{{\{\boldsymbol{\sigma}\}}}\bra{\{\boldsymbol{\sigma}\}}$, where
\begin{equation}
\label{toolbox}
\begin{split}
K_{\rm eff}(\{\boldsymbol{\sigma}\})&=\sum_{\alpha}\sum_{{\bf i}>{\bf j}}\tilde{J}_{{\rm d};{\bf i}{\bf j}}^{\alpha}(\sigma_{\bf i},\sigma_{\bf j})\ee^{ie^*\int_{{\bf j}}^{{\bf i}}{\rm d}{\bf r}\cdot {\bf A}^{\alpha}_{\rm s}(\sigma_{\bf i},{\sigma_{\bf j}})}b_{\alpha,{\bf i}}^{\dagger}b_{\alpha,{\bf j}}^{\phantom{\dagger}}\\
& +\sum_{\alpha}\sum_{{\bf i}}\textstyle{\frac{\epsilon_{\alpha}}{2}}\sigma_{\bf i}b_{\alpha,{\bf i}}^{\dagger}b_{\alpha,{\bf i}}^{\phantom{\dagger}}+\text{H.c.},\\
V_{\rm eff}&=\sum_{\ii,\alpha,\gamma}\tilde{U}_{\ii,\alpha\gamma}b_{\alpha,\ii}^{\dagger}b_{\gamma,\ii}^{\dagger}b_{\gamma,\ii}^{\phantom{\dagger}}b_{\alpha,\ii}^{\phantom{\dagger}},
\end{split}
\end{equation}
where the particular expression for the dressed tunneling and synthetic gauge field will depend on the configuration of the frequency gradient and the periodic driving. Let us list the possibilities that have been explored in this work:

{\it i) Dynamic localization:} The tunneling amplitude  $\tilde{J}_{{\rm d};{\bf i}{\bf j}}^{\alpha}=J_{{\rm t};{\bf i}{\bf j}}^{\alpha}\mathcal{F}_{0}(\eta_{\rm d},\eta_{\rm d},\Delta\phi_{{\bf i}{\bf j}})$ does not depend on the spin state, and the synthetic gauge field vanishes ${\bf A}^{\alpha}_{\rm s}=0$.  This is achieved in the regime of vanishing gradient $\Delta\omega_{\alpha}=0$, and setting the beatnote of the Raman beams $\omega_{\rm L}\ll \omega_{\alpha}$.   By tuning $\eta_{\rm d}$, one can find a value where the tunneling strength vanishes and thus the particles are dynamically localized, a phenomenon also known as coherent destruction of tunneling~\cite{pat_cm}.

{\it ii) Synthetic Abelian gauge fields:} The spin-independent tunneling amplitude is $\tilde{J}_{{\rm d};{\bf i}{\bf j}}^{\alpha}=J_{{\rm t};{\bf i}{\bf j}}^{\alpha}\mathcal{F}_{f({\bf i},{\bf j})}(\eta_{\rm d},\eta_{\rm d},\Delta\phi_{{\bf i}{\bf j}})$,  where   $f({\bf i},{\bf j})=r(i_1-j_1)$ depends on an integer $r$. The synthetic gauge potential is    ${\bf A}_{\rm s}({\bf r})=-B_0 y{\bf e}_1$, and follows from  the  regime with a finite gradient $\Delta\omega_{\alpha}\gg J_{{\rm t};{\bf i}{\bf j}}^{\alpha} $, such that the beatnote of the Raman laser beams fulfills the resonance condition  $\omega_{\rm L}=\Delta\omega_{\alpha}/r.$ In this case, phonons behave as charged particles that move in a two-dimensional plane pierced by an orthogonal magnetic field whose flux $\phi_2=\Delta k_2d_2$ can be modified by varying the  Raman wavevector. Phenomena typical of  integer quantum Hall samples~\cite{hofstadter_model,hofstadter_properties}, or bosonic flux ladders~\cite{flat_band_ladders}, can also be observed in this platform. Besides, in combination with strong phonon-phonon interactions, one can find bosonic versions of the fractional quantum Hall states~\cite{bhm_gauge}. 

{\it iii) Synthetic non-Abelian gauge fields:} The above scheme can be generalized to the non-Abelian gauge group SU(2), such that both in-plane vibrational modes play the role of a flavor component.  We have described in detail a particular SU(2) gauge field, which requires the vibrations along each direction to be subjected to an opposite frequency gradient. Using the same assumptions as in the Abelian case,  the tunneling amplitude becomes $\tilde{J}_{{\rm d};{\bf i}{\bf j}}^{\alpha}=J_{{\rm t};{\bf i}{\bf j}}^{\alpha}\mathcal{F}_{f_{\alpha}({\bf i},{\bf j})}(\eta_{\rm d\alpha},\eta_{\rm d\alpha},\Delta\phi_{{\bf i}{\bf j}})$, with $f_{\alpha}({\bf i},{\bf j})=r_{\alpha}(i_1-j_1)$ such that $r_x=-r_y=r$. Hence, the synthetic gauge field   ${\bf A}^{\rm na}_{\rm s}=-B_0y\tau_{z}{\bf e}_1$ becomes a SU(2) operator acting in the flavor space. This scheme opens a route towards a bosonic counterpart of the quantum spin Hall effect~\cite{qsh}.

{\it iv)  Bond and site disorder:} The dressed-tunneling amplitude $\tilde{J}_{{\rm d};{\bf i}{\bf j}}^{\alpha}(\sigma_{\bf i},\sigma_{\bf j})={J}_{{\rm t};{\bf i}{\bf j}}^{\alpha}\mathcal{F}_{f({\bf i},{\bf j})}(\eta_{\rm d}\sigma_{\ii},\eta_{\rm d}\sigma_{\bf j},\Delta\phi_{{\bf i}{\bf j}})$, and the on-site energies $\epsilon_{\alpha}\sigma_{\bf i}$,  take on different  values depending on the spin configuration. If the initial  state is a linear superposition of different spin configurations, the phonon dynamics is determined by a random Hamiltonian with bond and site disorder. This regime is achieved for a gradient $\Delta\omega_{\alpha}\gg J_{{\rm t};{\bf i}{\bf j}}^{\alpha} $, and laser beatnote  $\omega_{\rm L}=\Delta\omega_{\alpha}/r$ giving rise to a state-dependent periodic driving. Besides,  an additional Raman beatnote tuned  close to the atomic transition $\bar{\omega}_{\rm L}\approx\omega_0-\omega_{\alpha}$ gives raise to a phonon-dependent Stark shift  in the limit of large detuning. In the non-interacting regime, this tool allows us to explore the physics of Anderson localization~\cite{and_localization}.  By adding strong interactions, it yields gapless insulating phases such as the Bose glass~\cite{bhm}, the Mott glass~\cite{bhm_bond_disorder}, and the random-singlet glass~\cite{bhm_both_disorder}.

{\it vi) Decorated flux lattices:} In the absence of the frequency gradient, one can tune  the Raman lasers beatnote in resonance to the above phonon-dependent Stark shift $\omega_{\rm L}=2\epsilon_{\alpha}/r$. Once again, the dressed-tunneling amplitude becomes spin-dependent $\tilde{J}_{{\rm d};{\bf i}{\bf j}}^{\alpha}(\sigma_{\bf i},\sigma_{\bf j})=J_{{\rm t};{\bf i}{\bf j}}^{\alpha}\mathcal{F}_{f(\sigma_{\bf i},\sigma_{\bf j})}(\eta_{\rm d},\eta_{\rm d},\Delta\phi_{{\bf i}{\bf j}})$ where $f(\sigma_{\bf i},\sigma_{\bf j})=r(\sigma_{\bf i}-\sigma_{\bf j})/2$. Moreover, the synthetic gauge field also depends on the spin configuration and we can make decorated flux lattices as those shown in Fig.~\ref{fig_fluxes} by selecting a particular spin state. Phenomena related to  charged particles under inhomogeneous magnetic fields can be explored, such as staggered fields or $\pi$-flux lattices with defects.

\acknowledgments
 This work was partially supported by EU STREPs HIP,  PICC, QUITEMAD S2009-ESP-1594, FIS2009-10061, CAM-UCM/910758, and RyC Contract Y200200074.


\begin{references}

\bibitem{more}
P. W. Anderson, \href{http://www.sciencemag.org/content/177/4047/393.citation}{Science {\bf 177,} 393 (1972).}

\bibitem{feynman_lectures}
R. P. Feynman, {\it Statistical Mechanics: A Set Of Lectures} (Benjamin/Cummings Publishing, Massachusetts, 1972).

\bibitem{qs}
R.P.  Feynman, \href{http://www.springerlink.com/content/t2x8115127841630/}{Int. J. Theo. Phys. {\bf 21,} 467 (1982).}

\bibitem{blatt_wineland}
R. Blatt and D. Wineland, \href{http://www.nature.com/nature/journal/v453/n7198/abs/nature07125.html}{Nature {\bf 453,} 1008 (2008).}

\bibitem{hubbard_porras}
D. Porras and J.I. Cirac, \href{http://prl.aps.org/abstract/PRL/v93/i26/e263602}{Phys. Rev. Lett. {\bf 93,} 263602 (2004)}; X.-L. Deng, D. Porras and J.I. Cirac,
\href{http://pra.aps.org/abstract/PRA/v77/i3/e033403}{
Phys. Rev. A {\bf 77}, 03303 (2008).}

\bibitem{many_body_qs}
C. Schneider, D. Porras, and T. Schaetz, \href{http://arxiv.org/abs/1106.2597}{arXiv:1106.2597 (2011).}

\bibitem{gates}
J. I. Cirac and P. Zoller, \href{http://prl.aps.org/abstract/PRL/v74/i20/p4091_1}{Phys. Rev. Lett. {\bf 74,} 4091 (1995)}; C. Monroe, et al.,  \href{http://prl.aps.org/abstract/PRL/v75/i25/p4714_1}{Phys. Rev. Lett. {\bf 75,} 4714 (1995)}; F. Schmidt-
Kaler, et al. \href{http://www.nature.com/nature/journal/v422/n6930/full/nature01494.html}{Nature {\bf 422,} 408 (2003).} For a review, see~\cite{gate_review} and references therein.

\bibitem{gate_review}
K-A Brickman Soderberg and C. Monroe, \href{http://iopscience.iop.org/0034-4885/73/3/036401}{Rep. Prog. Phys. {\bf 73}  036401(2010). }


\bibitem{ions_ising_interaction}
J.I. Cirac, and P. Zoller, \href{http://www.nature.com/nature/journal/v404/n6778/full/404579a0.html}{Nature {\bf 404,} 579 (2000)}; G. J. Milburn, S. Schneider, and D. F. V. James,  \href{http://adsabs.harvard.edu//abs/2000ForPh..48..801M}{Fortschr. Phys. {\bf 48,} 801 (2000)}; F. Mintert and C. Wunderlich, \href{http://prl.aps.org/abstract/PRL/v87/i25/e257904}{Phys. Rev. Lett. {\bf 87,} 257904 (2001)}.

\bibitem{ising_porras}
D. Porras and J.I. Cirac, \href{http://prl.aps.org/abstract/PRL/v92/i20/e207901}{Phys. Rev. Lett. {\bf 92,} 207901 (2004).}

\bibitem{ising_exp}
A. Friedenauer, et {\it al.}, \href{http://www.nature.com/nphys/journal/v4/n10/abs/nphys1032.html}{Nature Phys. {\bf 4,} 757 (2008)}; K. Kim, et {\it al.}, \href{http://prl.aps.org/abstract/PRL/v103/i12/e120502}{Phys. Rev. Lett. {\bf 103,} 120502 (2009)}; K. Kim, et {\it al.} \href{http://www.nature.com/nature/journal/v465/n7298/full/nature09071.html}{Nature {\bf 465,} 590 (2010)};
E. E. Edwards, et {\it al.}, \href{http://prb.aps.org/abstract/PRB/v82/i6/e060412}{Phys. Rev. B {\bf 82,} 060412(R) (2010)}; R. Islam, et {\it al.},  \href{http://www.nature.com/ncomms/journal/v2/n7/pdf/ncomms1374.pdf?WT.ec_id=NCOMMS-201107}{Nat. Comm. {\bf 2,} 377 (2011).}

\bibitem{digital}
J. T. Barreiro et {\it al.}, \href{http://www.nature.com/nature/journal/v470/n7335/full/nature09801.html?WT.ec_id=NATURE-20110224}{Nature {\bf 470,} 486 (2011)}; B. P. Lanyon, et {\it al.} \href{http://www.sciencemag.org/content/early/2011/08/31/science.1208001.abstract}{Science {\bf 334,} 57  (2011).}

  \bibitem{neural_networks_lewenstein}
M. Pons, et {\it al.}, \href{http://prl.aps.org/abstract/PRL/v98/i2/e023003}{Phys. Rev. Lett. {\bf 98,} 023003 (2007).}

\bibitem{3_spin_models}
A. Bermudez, D. Porras, and M. A. Martin-Delgado, \href{http://pra.aps.org/abstract/PRA/v79/i6/e060303}{ Phys. Rev. A {\bf 79,} 060303 (R) (2009).}


\bibitem{frustrated}
P. Hauke, et {\it al.},\href{New J. Phys. 12, 113037 (2010)}{ New J. Phys. {\bf 12,}  113037 (2010).}; G.-D. Lin, C. Monroe, and L.-M. Duan, \href{http://prl.aps.org/abstract/PRL/v106/i23/e230402}{Phys. Rev. Lett. {\bf 106,} 230402 (2011).}; A. Bermudez, et {\it al.}, \href{http://prl.aps.org/abstract/PRL/v107/i20/e207209}{Phys. Rev. Lett. {\bf 107,} 207209 (2011).}

\bibitem{impurity_magnetism}
P. A. Ivanov and F. Schmidt-Kaler, \href{http://iopscience.iop.org/1367-2630/13/12/125008/article?genre=bookitem&sid=IOPP%3Ajnl_ref&spage=590&title=Nature&aufirst=K&volume=465&date=2010&v_showaffiliations=no&aulast=Kim}{ New J. Phys. {\bf 13,} 12500 (2011).}

\bibitem{top}
M. M\"{u}ller, et {\it al.}, \href{http://iopscience.iop.org/1367-2630/13/8/085007}{New J. Phys. {\bf 13,} 08500 (2011)}; R. Schmied, J. H. Wesenberg, and D. Leibfried, \href{http://iopscience.iop.org/1367-2630/13/11/115011/}{New J. Phys. {\bf 13,} 115011 (2011).}



 \bibitem{frustrated_hard_core_bosons_schmied}
R. Schmied, et {\it al.},  \href{http://iopscience.iop.org/1367-2630/10/4/045017}{New J. Phys. {\bf 10,} 045017 (2008).}

\bibitem{kontorova_garcia}
I. Garc\'ia-Mata, O. V. Zhirov, and D. L. Shepelyansky,  \href{http://www.springerlink.com/content/7329u86702003375/}{Eur. Phys. 
J. D {\bf 41,} 325 (2007)}; T. Pruttivarasin, et {\it al.}, \href{http://iopscience.iop.org/1367-2630/13/7/075012/}{New J. Phys. {\bf 13} 075012 (2011)}.

\bibitem{porras_spin_boson}
D. Porras, F. Marquardt, J. von Delft, and J.I. Cirac, \href{http://pra.aps.org/abstract/PRA/v78/i1/e010101}{Phys. Rev. A  {\bf 78,} 010101 (R) (2008).}


\bibitem{lattice_JC_ivanov}
P. A. Ivanov, et {\it al.}, \href{http://pra.aps.org/abstract/PRA/v80/i6/e060301}{Phys. Rev. A {\bf 80,} 060301(R) (2009).}

\bibitem{anderson}
A. Bermudez, M. A. Martin-Delgado, and D. Porras, \href{http://iopscience.iop.org/1367-2630/12/12/123016/pdf/1367-2630_12_12_123016.pdf}{ New J. Phys. {\bf 12,} 123016 (2010)}.


\bibitem{ion_dirac}
L. Lamata, J. Leon, T. Schaetz, and E. Solano, \href{http://prl.aps.org/abstract/PRL/v98/i25/e253005}{Phys. Rev. Lett. {\bf 98,} 253005 (2007)}; R. Gerritsma, et {\it al.},\href{http://www.nature.com/nature/journal/v463/n7277/abs/nature08688.html}{ Nature {\bf 463,} 68 (2010)}; R. Gerritsma, et {\it al.}, \href{http://prl.aps.org/abstract/PRL/v106/i6/e060503}{Phys. Rev. Lett. {\bf 106,} 060503, (2011).}




\bibitem{gauge}
A. Bermudez, T. Schaetz, and D. Porras, \href{http://prl.aps.org/abstract/PRL/v107/i15/e150501}{Phys. Rev. Lett. {\bf 107,} 150501 (2011).}



\bibitem{pat_cm}
D. H. Dunlap and V. M. Kenkre, \href{http://prb.aps.org/abstract/PRB/v34/i6/p3625_1}{Phys. Rev. B {\bf 34,} 3625 (1986)}; F. Grossman et {\it al.}, \href{http://prl.aps.org/abstract/PRL/v67/i4/p516_1}{Phys. Rev. Lett.  {\bf 67,} 516 (1991)}; M. Holthaus, \href{http://prl.aps.org/abstract/PRL/v69/i2/p351_1}{Phys. Rev. Lett. {\bf 69,} 351 (1992)}. 

\bibitem{pat_review}
 M. Grifoni and P. H\"{a}nngi, \href{http://www.sciencedirect.com/science/article/pii/S0370157398000222}{Phys. Rep. {\bf 304,} 229 (1998).}

 
 \bibitem{pat_ol}
 K. Drese and M. Holthaus, \href{http://www.sciencedirect.com/science/article/pii/S0301010497000256}{Chem. Phys. {\bf 217,} 201 (1997)};
 A. Eckardt et {\it al.}, \href{http://prl.aps.org/abstract/PRL/v95/i26/e260404}{Phys. Rev. Lett. {\bf 95,} 260404 (2005)}; C.E. Creffield et {\it al.}, \href{http://prl.aps.org/abstract/PRL/v96/i21/e210403}{Phys. Rev. Lett. {\bf 96,} 210403 (2006)}; H. Lignier, et {\it al.}, \href{http://prl.aps.org/abstract/PRL/v99/i22/e220403}{Phys. Rev. Lett. {\bf 99,} 220403 (2007)}; E. Kierig, et {\it al.},\href{http://prl.aps.org/abstract/PRL/v100/i19/e190405}{ Phys. Rev. Lett. {\bf 100,} 190405 (2008)}; A. Kolovsky, \href{http://iopscience.iop.org/0295-5075/93/2/20003/}{Europhys. Lett. {\bf 93,} 20003 (2011).}



\bibitem{inhibitted_hopping_exp}
K. R. Brown, et {\it al.}, \href{http://www.nature.com/nature/journal/v471/n7337/full/nature09721.html}{Nature {\bf 471,} 196 (2011)};
M. Harlander, et {\it al.}, \href{http://www.nature.com/nature/journal/v471/n7337/full/nature09800.html}{Nature {\bf 471,} 200 (2011)}.
\bibitem{microtraps}
T. Schaetz, et al., \href{http://www.tandfonline.com/doi/abs/10.1080/09500340701639631}{Jour. of Mod. Opt. {\bf 54,} 2317 (2007)}; J. Chiaverini and W. E. Lybarger, \href{http://pra.aps.org/abstract/PRA/v77/i2/e022324}{Phys. Rev. A {\bf 77,} 022324 (2008)}; J. Labaziewicz
et al., \href{http://prl.aps.org/abstract/PRL/v100/i1/e013001}{Phys. Rev. Lett. {\bf 100,} 013001 (2008).}

\bibitem{schmied_micro}
R. Schmied, J. H. Wesenberg, and D. Leibfried,\href{http://prl.aps.org/abstract/PRL/v102/i23/e233002}{ Phys. Rev. Lett. {\bf 102,} 233002 (2009)}.

\bibitem{photon_cavities}
S. Schmidt et al., \href{http://prb.aps.org/abstract/PRB/v82/i10/e100507}{Phys. Rev. B {\bf 82,} 100507(R) (2010)}.

\bibitem{pat_greiner}
R. Ma, et {\it al.}, \href{http://prl.aps.org/abstract/PRL/v107/i9/e095301}{Phys. Rev. Lett. {\bf 107,} 095301 (2011).}

\bibitem{laser_assisted_ol}
D. Jaksch and P. Zoller, \href{http://iopscience.iop.org/1367-2630/5/1/356}{New J. Phys. {\bf 5}, 56 (2003)}; see also J. Dalibard, F. Gerbier, G. Juzeli\"unas, and P. \"Ohberg, \href{http://rmp.aps.org/abstract/RMP/v83/i4/p1523_1}{Rev. Mod. Phys. {\bf 83,} 1523 (2011)}, and references therein,

\bibitem{ab_st}
M. Abramowitz and I. A. Stegun, {\it Handbook of Mathematical Functions,} (Dover Publications, New York, 1965)

\bibitem{ab_phase}
Y. Aharonov and D. Bohm,\href{http://prola.aps.org/abstract/PR/v115/i3/p485_1}{ Phys. Rev. {\bf 115,} 485 (1959)}.




\bibitem{lgt}
I. Montvay and G. Munster, {\it Quantum Fields on a Lattice,} (Cambridge University Press, New York, 1994). 

\bibitem{peierls}
R. Peierls, \href{http://www.springerlink.com/content/x2p66875h1j87374/}{Zeit. fur Physik {\bf 80}  763 (1933)}.

\bibitem{engineered_env}
C. J. Myatt, et {\it al.} \href{http://www.nature.com/nature/journal/v403/n6767/abs/403269a0.html}{Nature {\bf 403,} 269 (2000).}

\bibitem{hofstadter_model}
M.Ya. AzbelÕ, Zh. Eksp. Teor. Fiz. {\bf 46,} 929 (1964) [Sov. Phys.
JETP {\bf 19,} 634 (1964)]; P. G. Harper, \href{http://iopscience.iop.org/0370-1298/68/10/305}{Proc. Phys. Soc. A {\bf 68,}
874(1955)} ; D. R. Hofstadter, \href{Phys. Rev. B 14, 2239 (1976)}{Phys. Rev. B {\bf 14,} 2239 (1976).}

\bibitem{hofstadter_properties}
D. J. Thouless, M. Kohmoto, M. P. Nightingale, and M. den Nijs,\href{http://prl.aps.org/abstract/PRL/v49/i6/p405_1}{ Phys. Rev. Lett. {\bf 49,}  405 (1982)}; D. J. Thouless and Q. Niu, \href{http://iopscience.iop.org/0305-4470/16/9/015}{J. Phys. A: Math. Gen.  {\bf 16,} 1911 (1983)}; R. Rammal, G. Toulouse, M. T. Jaekel, and B. I. Halperin, \href{http://prb.aps.org/abstract/PRB/v27/i8/p5142_1}{Phys. Rev. B {\bf 27,} 5142 (1983)}; E. H. Lieb, \href{http://prl.aps.org/abstract/PRL/v73/i16/p2158_1}{Phys. Rev. Lett. {\bf 73,}  2158 (1994)}.

\bibitem{flat_band_ladders}
J. Vidal, R. Mosseri, and B. Doucot,\href{http://prl.aps.org/abstract/PRL/v81/i26/p5888_1}{ Phys. Rev. Lett. {\bf 81,} 5888 (1998)}; J. Vidal, B. Doucot , R. Mosseri, and P. Butaud, \href{http://prl.aps.org/abstract/PRL/v85/i18/p3906_1}{Phys.
Rev. Lett. {\bf 85,} 3906 (2000)}; M. Creutz, \href{http://prl.aps.org/abstract/PRL/v83/i13/p2636_1}{Phys. Rev. Lett. {\bf 83,} 2636 (1999)}; A. Bermudez, D. Patane, L. Amico, and M.A. Martin-Delgado, \href{http://prl.aps.org/abstract/PRL/v102/i13/e135702}{Phys. Rev. Lett. {\bf 102,} 135702 (2009)}.

\bibitem{bhm}
M.P.A. Fisher, et {\it al.}, \href{http://prb.aps.org/abstract/PRB/v40/i1/p546_1}{Phys. Rev. B {\bf 40,} 546 (1989).}

\bibitem{bhm_gauge}
A. S. Sorensen, E. Demler, and M. D. Lukin, \href{http://prl.aps.org/abstract/PRL/v94/i8/e086803}{Phys. Rev. Lett. {\bf 94,} 086803 (2005)}; M. Hafezi, et {\it al.}, \href{http://pra.aps.org/abstract/PRA/v76/i2/e023613}{Phys. Rev. A {\bf 76,} 023613 (2007)}; G. M\"oller and N. R. Cooper, \href{http://prl.aps.org/abstract/PRL/v103/i10/e105303}{Phys. Rev. Lett. {\bf 103,} 105303 (2009).}
\bibitem{bhm_atoms}
D. Jaksch, et {\it al.}, \href{http://prl.aps.org/abstract/PRL/v81/i15/p3108_1}{Phys. Rev. Lett. {\bf 81,}  3108 (1998)}; M. Greiner, et {\it al.}, \href{http://www.nature.com/nature/journal/v415/n6867/abs/415039a.html}{Nature {\bf 415,}  39 (2002).}



\bibitem{ti_review}
M. Z. Hasan and C. L. Kane, \href{http://rmp.aps.org/abstract/RMP/v82/i4/p3045_1}{Rev. Mod. Phys. {\bf 82,} 3045 (2010)}; X. L. Qi and S. C. Zhang, \href{http://rmp.aps.org/abstract/RMP/v83/i4/p1057_1}{	Rev. Mod. Phys. {\bf 83,} 1057 (2011).}

\bibitem{ion_trap_reviews}
D. J. Wineland, et al., \href{http://arxiv.org/abs/quant-ph/9710025}{J. Res. Natl. Inst. Stand. Tech. {\bf 103,} 259 (1998)};
D. Leibfried, et al., \href{http://rmp.aps.org/abstract/RMP/v75/i1/p281_1}{Rev. Mod. Phys. {\bf 75,} 281 (2003)}; H. Haeffner, et al., \href{http://www.sciencedirect.com/science/article/pii/S0370157308003463}{Phys. Rep. {\bf 469,} 155 (2008)}.

\bibitem{james}
D. F. V. James, \href{http://www.springerlink.com/content/01e0nug5vlv9dta2/}{Appl. Phys. B {\bf 66,} 181 (1998)}.

\bibitem{non_abelian}
A. Zee, {\it Quantum Field Theory in a Nutshell}, (Princeton University Press, 2003).

\bibitem{qsh}
C. L. Kane and E. J. Mele, \href{http://prl.aps.org/abstract/PRL/v95/i14/e146802}{Phys. Rev. Lett. {\bf 95,} 146802 (2005)}; B. A. Bernevig, T. L. Hughes and S. C. Zhang, \href{http://www.sciencemag.org/content/314/5806/1757.abstract}{Science {\bf 314,}
1757 (2006)}; N. Goldman, et {\it al.}, \href{http://prl.aps.org/abstract/PRL/v105/i25/e255302}{Phys. Rev. Lett. {\bf 105,} 255302 (2010)}.




\bibitem{ziman_disorder}
J.M. Ziman, {\it Models of Disorder: The Theoretical Physics of Homogeneously Disordered Systems}, (Cambridge University Press, 1979.)

\bibitem{qs_random}
U. Gavish and Y. Castin, \href{http://prl.aps.org/abstract/PRL/v95/i2/e020401}{Phys. Rev. Lett. {\bf 95,} 020401 (2005)}; B. Paredes, F. Verstraete, and J.I. Cirac, \href{http://prl.aps.org/abstract/PRL/v95/i14/e140501}{Phys. Rev. Lett. {\bf 95,}140501 (2005)}.

\bibitem{bhm_bond_disorder}
E. Altman, Y. Kafri, A. Polkovnikov, and G. Refael,  \href{http://prl.aps.org/abstract/PRL/v93/i15/e150402}{Phys. Rev. Lett. {\bf 93,} 150402 (2004)}; K. G. Balabanyan, N. ProkofÕev, and B. Svistunov, \href{http://prl.aps.org/abstract/PRL/v95/i5/e055701}{Phys. Rev. Lett. {\bf 95,} 055701 (2005)}; P. Sengupta and S. Haas, \href{http://prl.aps.org/abstract/PRL/v99/i5/e050403}{Phys. Rev. Lett. {\bf 99,} 050403 (2007).}


\bibitem{and_localization}
 P. W. Anderson, \href{http://prola.aps.org/abstract/PR/v109/i5/p1492_1}{Phys. Rev. {\bf 109,} 1492 (1958)};  B. Kramer and A. MacKinnon, \href{http://iopscience.iop.org/0034-4885/56/12/001}{Rep. Prog. Phys. {\bf 56,} 1469
(1993).}

\bibitem{bhm_both_disorder}
E. Altman, Y. Kafri, A. Polkovnikov, and G. Refael, \href{http://prl.aps.org/abstract/PRL/v100/i17/e170402}{Phys. Rev. Lett. 100, 170402 (2008)}.

\bibitem{anyons_flux}
C. Weeks, G. Rosenberg, B. Seradjeh, and M. Franz, \href{http://www.nature.com/nphys/journal/v3/n11/abs/nphys730.html}{Nature Phys. {\bf 3,} 796 (2007)}; C. Rosenberg, B. Seradjeh, C. Weeks, and M. Franz, \href{http://prb.aps.org/abstract/PRB/v79/i20/e205102}{Phys. Rev. B {\bf 79,} 205102 (2009)}.

\end{references}
\end{document}